\documentclass{aa}
\usepackage{natbib}
\bibpunct{(}{)}{;}{a}{}{,} 
\usepackage{graphicx}
\usepackage{aalongtable}
\usepackage{lscape}
\usepackage{txfonts}
\begin{document}
\title{The observed spiral structure of the Milky Way\thanks{Appendix
    A is available in electronic form at http$://$www.aanda.org.}
  \thanks{Full Tables A.1–A.3 are only available at the CDS via
    anonymous ftp to cdsarc.u-strasbg.fr (130.79.128.5) or via
    http$://$cdsarc.u-strasbg.fr/viz-bin/qcat?J/A+A/569/A125 and also
    at the authors’ webpage: http$://$zmtt.bao.ac.cn/milkyway/}}
\author{L.~G. Hou \inst{}
    \and J.~L. Han
    \inst{}
  }


  \institute{National Astronomical Observatories, Chinese Academy of
    Sciences,
    Jia-20, DaTun Road, ChaoYang District, 100012 Beijing, PR China\\
    \email{lghou@nao.cas.cn, hjl@nao.cas.cn} }

   \date{Received 21 April 2014; accepted 7 July 2014}


  \abstract
  {The spiral structure of the Milky Way is not yet well
    determined. The keys to understanding this structure are to
    increase the number of reliable spiral tracers and to determine
    their distances as accurately as possible. HII regions, giant
    molecular clouds (GMCs), and 6.7 GHz methanol masers are closely
    related to high mass star formation, and hence they are excellent
    spiral tracers. The distances for many of them have been
    determined in the literature with trigonometric, photometric
    and/or kinematic methods.}
%
%
{We update the catalogs of Galactic HII regions, GMCs, and 6.7 GHz
  methanol masers, and then outline the spiral structure of the Milky
  Way.}
%
%
  {We collected data for more than 2500 known HII regions, 1300 GMCs,
    and 900 6.7 GHz methanol masers. If the photometric or
    trigonometric distance was not yet available, we determined the
    kinematic distance using a Galaxy rotation curve with the current
    IAU standard, $R_0$ = 8.5~kpc and $\Theta_0$ = 220 km~s$^{-1}$,
    and the most recent updated values of $R_0$ = 8.3~kpc and
    $\Theta_0$ = 239 km~s$^{-1}$, after velocities of tracers are
    modified with the adopted solar motions. With the weight factors
    based on the excitation parameters of HII regions or the masses of
    GMCs, we get the distributions of these spiral tracers.  }
%
%
  {The distribution of tracers shows at least four segments of arms in
    the first Galactic quadrant, and three segments in the fourth
    quadrant. The Perseus Arm and the Local Arm are also delineated by
    many bright HII regions. The arm segments traced by massive star
    forming regions and GMCs are able to match the HI arms in the
    outer Galaxy. We found that the models of three-arm and four-arm
    logarithmic spirals are able to connect most spiral tracers. A
    model of polynomial-logarithmic spirals is also proposed, which
    not only delineates the tracer distribution, but also matches the
    observed tangential directions.}
%
  {}


\maketitle
%

\section{Introduction}
\label{sect:intro}

The Milky Way is known to be a barred spiral galaxy, as shown by
surveys of molecular gas \citep[][]{dht01}, neutral gas
\citep[][]{burt88}, and by infrared star counts \citep[][]{bcb05}. We
do not know its precise structure, e.g., the number and position of
spiral arms, the location and length of the Galactic bar, however. The
obstacle is that the Sun is located on the edge of the Galactic disk
close to the Galactic plane, so that we can only see the superposition
of various structure features (e.g., spiral arms, spurs, arm branches,
and Galactic bars) along the observed line of sight. It is difficult
to disentangle the features and determine the relative position of
spiral tracers accurately.

To outline the precise structure of the Galaxy, a large number of
spiral tracers should be detected over the disk with distances
determined as accurately as possible. Observationally, it has been
shown that the narrow and sharply defined spiral arms of a galaxy
could be well delineated by the distribution of massive star forming
regions or giant molecular clouds \citep[GMCs;
  e.g.,][]{iga75,gg76,dwbw80,gcbt88,ggc93,rus03,bt08,hhs09}. HII
regions and 6.7 GHz methanol masers are situated in massive star
forming regions, and closely related to massive young stars. GMCs are
the birthplace of massive stars, and have been widely used to reveal
bright spiral arms and the gas arms of galaxies \citep[][]{bt08}. In
the Milky Way, they are particularly useful in mapping the spiral
pattern, because they are bright and can be widely detected throughout
the Galactic plane \citep[as far as $\sim$20 kpc away from the Sun,
  e.g.,][]{abbr11,dt11}. The radio recombination lines from HII
regions, the methanol maser lines near 6.7 GHz, and the molecular
rotational transitions are almost unabsorbed by the interstellar
medium.

Much effort has been dedicated to search for Galactic HII regions,
GMCs, or masers, then to determine their distances to infer the spiral
structure of our Galaxy. A good amount of observational data exist in
literature
\citep[e.g.,][]{dwbw80,ch87,srby87,gcbt88,ahck02,rus03,pdd04,swa+04}. We
have cataloged from the literature 815 HII regions and 963 GMCs with
known trigonometric, photometric, or kinematic distances, and combined
all these data to delineate the spiral structure of the Milky Way
\citep{hhs09}. A polynomial-logarithmic model of spiral arms was
proposed to fit the data distribution and match the observed
tangential directions of spiral arms.

In the last few years, much progress toward understanding the spiral
structure of our Galaxy has been made. First of all, a large number of
new spiral tracers have been detected
\citep[e.g.,][]{rjj+09,abbr11,gm11,gbnd14}. Some newly discovered HII
regions or GMCs even as far as $\sim$20 kpc from the Sun
\citep[][]{abbr11,dt11} provide important data to reveal the spiral
structure in the outer Galaxy. The kinematic distance ambiguity for
many HII regions, GMCs, and 6.7 GHz methanol masers has been solved
\citep[e.g.,][]{rjh+09,lra+11,gm11,abbr12,uhl12,jd12}, which enables
us to estimate their kinematic distances. In addition, the more
reliable trigonometric or photometric distances of a number of HII
regions or masers have been measured
\citep[e.g.,][]{rbr+10,shr+10,mdf+11,xmr+11,xlr+13,rza+12,zms+14,fb14}. The
available spiral tracers have almost {\it doubled} as compared to
those listed in \citet{hhs09}.

Second, some fundamental parameters of the Galaxy have been
observationally modified. Based on the stellar kinematics in the solar
neighborhood, \citet{sbd10} revised the solar motions with respect to
the Local Standard of Rest (LSR) as $U_\odot=11.10$~km~s$^{-1}$
(radially inward), $V_\odot=12.24$~km~s$^{-1}$ (in the direction of
Galactic rotation), and $W_\odot=7.25$~km~s$^{-1}$ (vertically upward,
all in J2000), which are inherently incorporated in the LSR velocities
of spiral tracers, hence influence the estimation of their kinematic
distances. A consistent value of the distance of the Sun to the
Galactic center (GC), $R_0=8.3$~kpc, was proposed by \citet{brm+11},
who incorporated studies on stellar orbits in the GC
\citep[][]{gsw+08,get+09} and trigonometric parallax measurements
toward Sgr\,B2~\citep{rmz+09} and massive star forming regions
\citep{rmz+09b,brm+11}. The circular rotation of the LSR, $\Theta_0$,
was estimated to be 239~km~s$^{-1}$ \citep[][]{brm+11}. Some recent
papers \citep[e.g.,][]{scho12,rmb+14} confirmed the results of
\citet{brm+11}. The values of both $R_0$ and $\Theta_0$ are different
from the IAU recommended values, $R_0=$~8.5~kpc and
$\Theta_0=$~220~km~s$^{-1}$.

The exciting new progress mentioned above motivates us to update the
catalogs of the spiral tracers in \citet{hhs09}, namely, the HII
regions, GMCs, and also to add the 6.7 GHz methanol masers in this
paper, and then to delineate the spiral structure of the Milky
Way. This work is organized as follows. In Sect. 2, we update the
catalogs of spiral tracers and make them publicly available. In
Sect. 3, we discuss the distributions of different kinds of spiral
tracers and their combination. Then, we use the models of three- and
four-logarithmic spiral arms and a model of polynomial-logarithmic
spirals to connect the identified arm segments. Discussions and
conclusions are given in Sect. 4.


\section{Tracers for the Galactic spiral structure}
\label{sect:Obs}

\begin{figure}
\centering\includegraphics[width=0.47\textwidth]{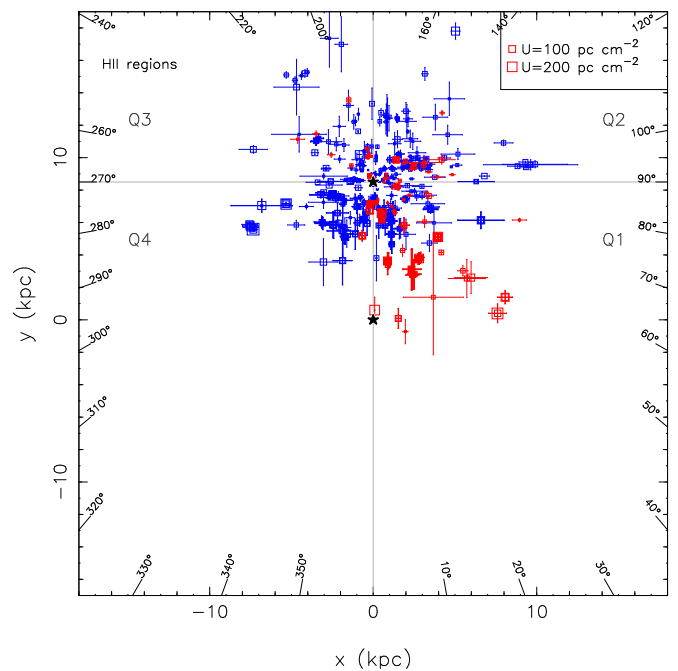}
\caption{Distribution of HII regions with photometric (blue) or
  trigonometric (red) distances. Two black stars indicate the location
  of the Sun ($x=0.0$ kpc, $y=8.5$ kpc) and the GC ($x=0.0$ kpc,
  $y=0.0$ kpc). Q1 $-$ Q4 indicate the four Galactic
  quadrants. Position uncertainties are indicated by error bars in $X$
  and $Y$ directions.}
\label{dis_op_all}
\end{figure}

The primary tracers for spiral arms are HII regions, GMCs, and 6.7 GHz
methanol masers. To outline a precise spiral pattern, the distances of
spiral tracers should be determined as accurately as possible. At
present, the most reliable measurements come from the trigonometric
parallax observations \citep[e.g.,][]{xrzm06,hna+12,rmb+14} and the
photometric method \citep[e.g.,][]{rag07,fb14}. The HII regions with
known photometric or trigonometric distances are shown in
Fig.~\ref{dis_op_all}, which cannot be used to infer the large-scale
spiral pattern of our Galaxy because data are sparse and limited
within about 6~kpc from the Sun.

To uncover the spiral pattern of the Milky Way, we have to involve the
large number of spiral tracers with kinematic distances at
present. The spiral tracers in the inner Galaxy (i.e., the distances
to the GC less than $R_0$) have two possible kinematic distances
corresponding to one $V_{\rm LSR}$, which is known as the kinematic
distance ambiguity. The ambiguity of a HII region can be solved by the
HI/H$_2$CO emission/absorption
observations~\citep[e.g.,][]{was+03,swa+04,ab09} and by the HI
self-absorption method~\citep[e.g.,][]{tl08,ab09}. The latter was also
applied to solve the distance ambiguity for
GMCs~\citep[e.g.,][]{rjh+09} and 6.7 GHz methanol
masers~\citep[e.g.,][]{gm11}.

\subsection{Tracer data}

\subsubsection{ HII regions:}

Based on the distribution of $\sim$100 complexes of H$\alpha$ and/or
H109$\alpha$ emission sources, i.e., bright HII regions,
\citet[][hereafter GG76]{gg76} first proposed a Galactic spiral
pattern with four arm segments. \citet{dwbw80} and \citet{ch87}
discovered more HII regions in the first and fourth Galactic
quadrants, which improved the model of GG76. \citet{ahck02},
\citet{was+03}, and \citet{swa+04} observed H110$\alpha$ and H$_2$CO
4.8 GHz lines simultaneously toward many ultracompact HII (UCHII)
regions, and resolved their kinematic distance ambiguity. The
distribution of UCHII regions traces three arm segments in the first
Galactic quadrant. \citet{rus03} compiled a large sample of $\sim$400
star forming complexes, and fitted the data with models of
two-,~three- and four-logarithmic spirals. She found that the four-arm
model is more appropriate to represent the data
distribution. \citet{pdd04} collected $\sim$180 HII regions with
known distances and found that the distribution is consistent with the
model of GG76.

In the past few years, there have been many HII region
measurements. Notably, Green Bank Telescope HII Region Discovery
Survey\footnote{http$://$www.cv.nrao.edu/hrds/}\citep[][]{babr10,abbr11}
and its extension \citep{abb+13} detected 603 discrete hydrogen radio
recombination line components toward 448 targets. About 60 of them
have negative velocities in the first Galactic quadrant or have
positive velocities in the fourth Galactic quadrant, implying that
they are located in the outer Milky Way or in the 3 kpc Arms
\citep[][]{dt08}. The kinematic distance ambiguity for the 149 newly
discovered HII regions was solved by \citet{abbr12}. The Arecibo HII
Region Discovery Survey \citep[][]{bab12} has detected 37 new HII
regions, kinematic distances were derived for 23 of
them. \citet{dze+11} and \citet{hze+11} observed H110$\alpha$ and
H$_2$CO 4.8 GHz lines simultaneously toward more than 200 candidates
of UCHII regions with the Urumqi 25~m telescope, and obtained
kinematic distances for about 20 HII regions. \citet{lra+11} solved
the kinematic distance ambiguity problem for more than 800 CS~(2$-$1)
emission sources~\citep[][]{bnm96}, which are associated with the
confident candidates of UCHII regions \citep[][]{wc89a}. By analyzing
the HI absorption data toward a sample of compact HII regions
identified from the Red MSX Source Survey \citep{hlo+05},
\citet{uhl12} solved the kinematic distance ambiguity for about 105
HII regions. Similar work was conducted by \citet{jd12} for 75 HII
regions. In addition, the relative accurate photometric or
trigonometric distances have been measured or revised for some HII
regions. \citet{pco10} identified the ionizing stars in nine HII
regions, and derived their photometric distances. \citet{mdf+11} used
the near-infrared color images to study the stellar content of 35 HII
regions, and estimated their spectrophotometric distances.
\citet{rza+12} analyzed the data of OB stars in two star forming
complexes \citep[see][]{rus03}, NGC\,6334 and NGC\,6357, and revised
their photometric distances. \citet{fb14} obtained the
spectrophotometric distances for 103 HII regions in the second and
third Galactic quadrants. By measuring the trigonometric parallax of
associated masers
\citep[e.g.,][]{rbr+10,okh+10,shr+10,nnh+11,shd+11,srd+12,irm+13,xlr+13,zrm+13,wsr+14},
the distances of about 30 HII regions have been determined with high
accuracy\footnote{The Bar and Spiral Structure Legacy (BeSSeL)
  Survey. http$://$bessel.vlbi-astrometry.org}\fnmsep\footnote{The
  Japanese VLBI Exploration of Radio Astrometry
  (VERA). http$://$veraserver.mtk.nao.ac.jp}.

We collect the HII regions with line measurements from the
references. The catalog contains about 4500 line measurements toward
more than 2500 HII regions or confident candidates. More than 1800 of
them have their distances determined. The photometric or trigonometric
distance and the solutions of the kinematic distance ambiguity if
available are collected from literature and listed in
Table~\ref{tab_a1}.

\subsubsection{GMCs}

Giant molecular clouds have proven to be good tracers of the spiral
structure \citep[][]{cdt86,dect86,gcbt88}. The famous
Sagittarius-Carina Arm was clearly delineated by the distribution of
GMCs \citep{gcbt88}. \citet{srby87} identified 268 GMCs in the first
Galactic quadrant and outlined three arm segments. \citet{efre98}
collected the known GMCs and fitted the data with a four-arm
model. \citet{rjh+09} solved the kinematic distance ambiguity of 750
GMCs identified from the Galactic Ring Survey of $^{13}$CO (1$-$0)
\citep[][]{jrs+09,rjj+09}, and showed three arm segments in the
distribution of GMCs. \citet{dt11} identified ten $^{12}$CO (1$-$0)
emission components having negative $V_{\rm LSR}$ in the first
Galactic quadrant, which are probably located in a spiral arm lying
outside the Outer Arm. For the fourth Galactic quadrants, a
preliminary identification of GMCs in the inner Galaxy was made by
\citet{bnt89}. Recently, \citet{gbnd14} reanalyzed the Columbia $-$
U. de Chile Survey data of CO, and identified 92 GMCs. The kinematic
distance ambiguity for 87 of them has been solved.

The GMC catalog in \citet{hhs09} is now updated to include these new
results. Molecular clouds with a mass greater than 10$^4$ $M_\odot$
are regarded to be massive enough and considered as GMCs in this
work. We collect about 1300 GMCs as listed in Table~\ref{tab_a2}, and
more than 1200 of them have distances determined.

\begin{figure*}[t]
\centering\includegraphics[width=170mm]{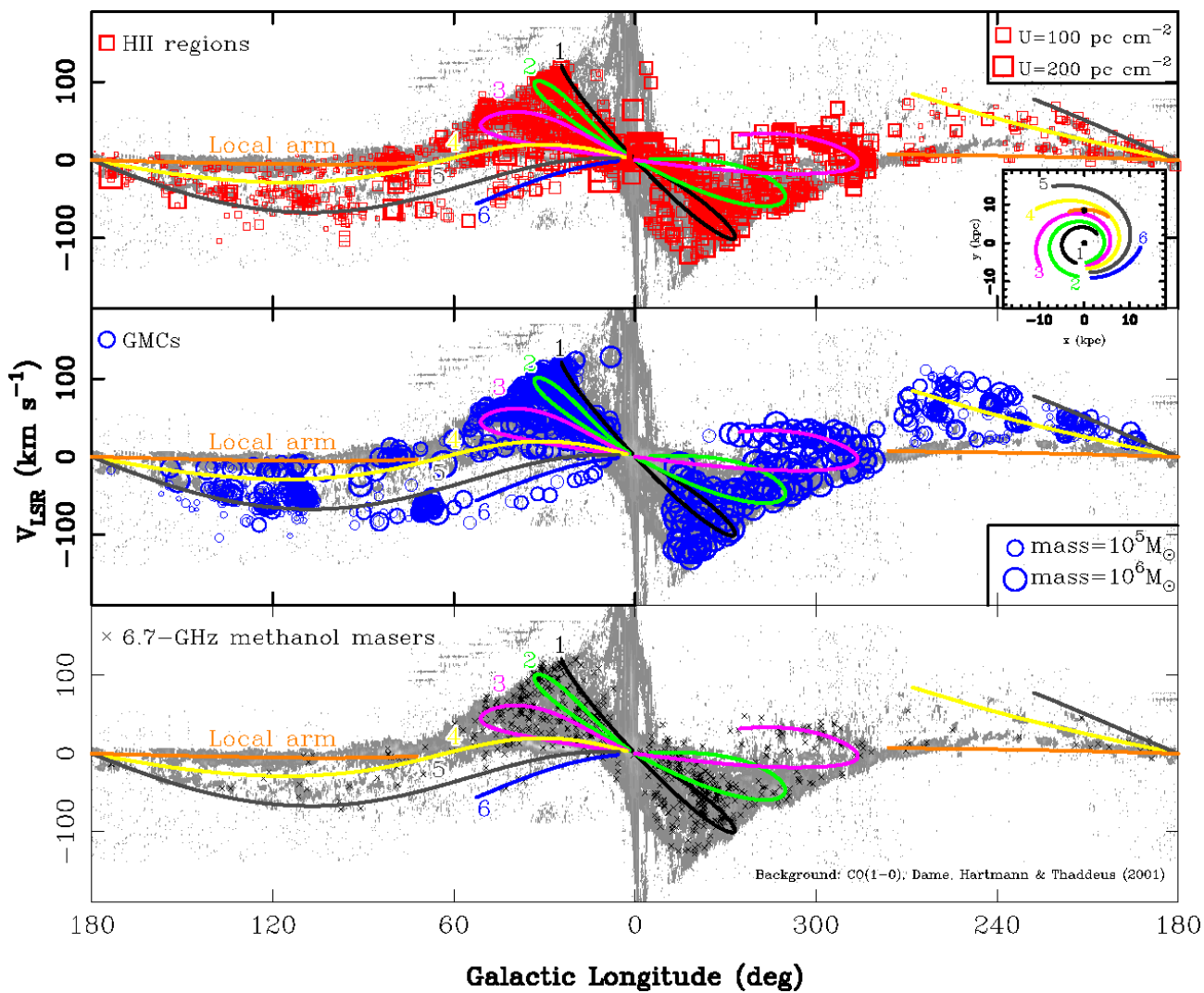}
\caption{Longitude-velocity diagrams for the HII regions ({\it
    upper}), GMCs ({\it middle}), and 6.7 GHz methanol masers ({\it
    lower}), which are overlaid on the image of $^{12}$CO~(1$-$0) of
  \citet[][the IAU standard solar motions are adopted]{dht01}.  The
  open squares indicate HII regions with the symbol area proportional
  to the excitation parameters (see Eq.~\ref{eq_U}). The open circles
  indicate the GMCs with the symbol size proportional to
  log($M_{GMCs}$) (see Sect. 2.4). The black crosses indicate the
  6.7 GHz methanol masers with a constant symbol size. A spiral arm
  model (see the left panel of Fig.~\ref{model_pol}) is overlaid to
  indicate the locations of the identified arm segments: arm-1: the
  Norma Arm; arm-2: the Scutum-Centaurus Arm; arm-3: the
  Sagittarius-Carina Arm; arm-4: the Perseus Arm; arm-5: the Outer
  Arm; arm-6: the Outer+1 Arm; and the Local Arm is also given. }
\label{lvall}
\end{figure*}

\subsubsection{6.7 GHz methanol Masers}

The 6.7 GHz methanol masers \citep{ment91} have been widely detected
from massive star forming regions in our Galaxy
\citep[e.g.,][]{pmb05,elli06,gcf+10,gcf+12}. Early detections of 519
6.7 GHz methanol masers were summarized by \citet{pmb05}. Since then,
\citet{pgd07} detected 86 masers in a 18.2 deg$^2$ area with the
Arecibo, and 48 of them are new detections. \citet{elli07} observed
200 {\it Spitzer}/GLIMPSE sources with Mt Pleasant 26~m and Ceduna
30~m radio telescopes, and they detected methanol masers in 38 sites,
nine of which are new. \citet{xlh+08} searched for 6.7 GHz methanol
masers toward a sample of 22 GHz water maser sources with the
Effelsberg 100~m telescope, and got ten new detections. \citet{cbhc09}
measured 18 methanol masers toward a sample of young stellar objects
with the VLA. The most important progress of the 6.7 GHz methanol
maser discovery is the Methanol Multibeam (MMB) Survey of the entire
Galactic plane
\citep[e.g.,][]{gcf+09,gcf+10,gcf+12,cfg+10,cfg+11}. \citet{fcf10}
searched for the 6.7 GHz methanol masers toward 296 massive star
forming regions, and detected 55 masers, 12 of which are
new. \citet{oah+13} performed a sensitive search of methanol masers
toward 107 sources discovered in the {\it Herschel} Infrared Galactic Plane
Survey \citep{msb+10}, and detected 32 masers, 22 of which are
new. \citet{sxc+14} searched for 6.7 GHz methanol masers toward 318
dust clumps, and found 29 masers, 12 of which are new. In addition,
the distances for many 6.7 GHz methanol masers have been
determined. \citet{pmg09} solved the kinematic distance ambiguity for
86 masers with the HI self-absorption method. \citet{gm11} collected a
large sample of 6.7 GHz methanol masers from references and new
detections from the MMB survey. Their database covers the longitude
ranges of 270$^\circ \leq$ $l$ $\leq$ 358$^\circ$, 5$^\circ \leq$ $l$
$\leq$ 67$^\circ$, and the latitude range of $|b| \leq$
1.5$^\circ$. The kinematic distance ambiguity for more than 400
methanol masers was solved by the HI self-absorption method.

We collect more than 900 methanol masers from the references above
(see Table~\ref{tab_a3}), about 750 of which have distances
determined, mostly kinematic, though some were measured by the
trigonometric parallax or the photometric method. The observed spectra
of 6.7 GHz methanol masers toward massive star forming regions always
have multicomponents. The velocities of the maser spots toward a
source are in a range of typically a few km~s$^{-1}$, sometimes more
than 10 km~s$^{-1}$~\citep[e.g., see][]{gcf+10}. The median velocity
instead of the peak velocity was suggested to be a better proxy of the
systematic velocity of the source~\citep{gm11}. Therefore, in our work,
we use the median velocities to calculate the kinematic distances of
masers. The systematic velocities and hence the distances determined
from the observed maser spectra may be affected by noncircular motions
and therefore have large uncertainties.

\subsubsection{Longitude-velocity diagram of spiral tracers}

The gas component in the Milky Way has been systematically observed
with the CO rotational transitions or the HI 21cm line. The
longitude-velocity diagrams represent the kinematic and dynamic
features of interstellar gas, which offer observational constraints on
the spiral structure \citep[e.g.,][]{dht01}. The key step in {\it
  mapping} the gaseous spiral arms is to reconstruct the gas
distribution from the observed longitude-velocity maps. In the inner
Galaxy, the distances of gas components are difficult to
determine. The spiral structure derived from diffuse HI and CO
distributions is far from clear at present \citep[e.g.,
  see][]{ns03,ns06}.

In contrast, HII regions, GMCs, and masers serve as excellent tracers
of spiral arms. Their distances can be obtained by trigonometric,
photometric, or kinematic method. As shown in Fig.~\ref{lvall},
almost all features in the longitude-velocity map of $^{12}$CO~(1$-$0)
\citep{dht01}\footnote{http$://$www.cfa.harvard.edu/mmw/MilkyWayinMolClouds.html}
have counterparts in the distributions of collected HII regions and
masers. With distances, one can reveal the unknown spiral arms.

\subsection{Fundamental parameter $R_0$, $\Theta_0$, and solar motions}

For most of the spiral tracers without photometric or trigonometric
distances, we have to obtain their kinematic distances, which depend
on the adopted rotation curve and the fundamental parameters of the
Galaxy, including the distance to the GC, $R_0$, the circular orbital
speed at the Sun, $\Theta_0$, and the solar motions with respect to
the LSR.

According to the IAU standard, the distance to the GC is
$R_0=8.5$~kpc, and the circular orbital speed at the Sun is
$\Theta_0=220$~km~s$^{-1}$. By measuring the S-star orbits in the GC
regions, \citet{gsw+08} obtained $R_0=8.4\pm0.4$~kpc, and
\citet{get+09} estimated $R_0=8.33\pm0.35$~kpc. By measuring the
trigonometric parallax toward Sgr\,B2, \citet{rmz+09} obtained a value
of $R_0=7.9^{+0.8}_{-0.7}$~kpc. \citet{rmz+09b} and \citet{brm+11}
fitted the parallax measurements of massive star forming regions and
derived the best fitted value of $R_0=8.4\pm0.6$~kpc and
$\Theta_0=254\pm16$~km~s$^{-1}$. A weighted average of
$R_0=8.3\pm0.23$~kpc was derived from the above measurements by
\citet{brm+11}, and the fitted value of $\Theta_0$ is in the range of
223$-$280~km~s$^{-1}$ for different rotation curves. The ratio of
$\Theta_0/R_0$ has been well constrained to be
$29.4\pm0.9$~km~s$^{-1}$~kpc$^{-1}$. \citet{brm+11} suggested a value
of $\Theta_0=239\pm7$~km~s$^{-1}$. \citet{scho12} obtained
$R_0=8.27\pm0.29$~kpc and $\Theta_0=238\pm9$~km~s$^{-1}$ by analyzing
the stellar kinematics. \citet{hna+12} found $R_0=8.05\pm0.45$~kpc and
$\Theta_0=238\pm14$~km~s$^{-1}$, using the VLBI astrometry data of 52
Galactic masers. \citet{rmb+14} estimated that $R_0=8.34\pm0.16$~kpc
and $\Theta_0=240\pm8$~km~s$^{-1}$ by analyzing the new parallax
measurements toward massive star forming regions. \citet{cfg+14} found
$R_0=8.36\pm0.11$~kpc with a statistical study on star cluster
dynamics and S-star orbits.

\begin{figure}[!b]
  \centering\includegraphics[width=0.47\textwidth]{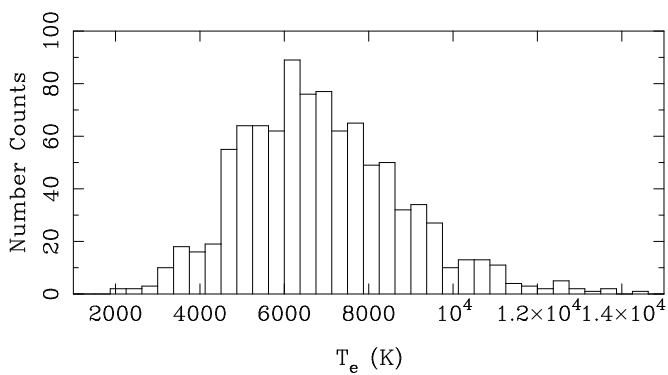}
  \caption{Distribution of estimated electron temperature ($T_e$) of
    HII regions from radio recombination line and radio continuum
    observations in Table~\ref{tab_a1}.}
\label{dis_te}
\end{figure}

\begin{figure}[!t]
\centering\includegraphics[width=0.47\textwidth]{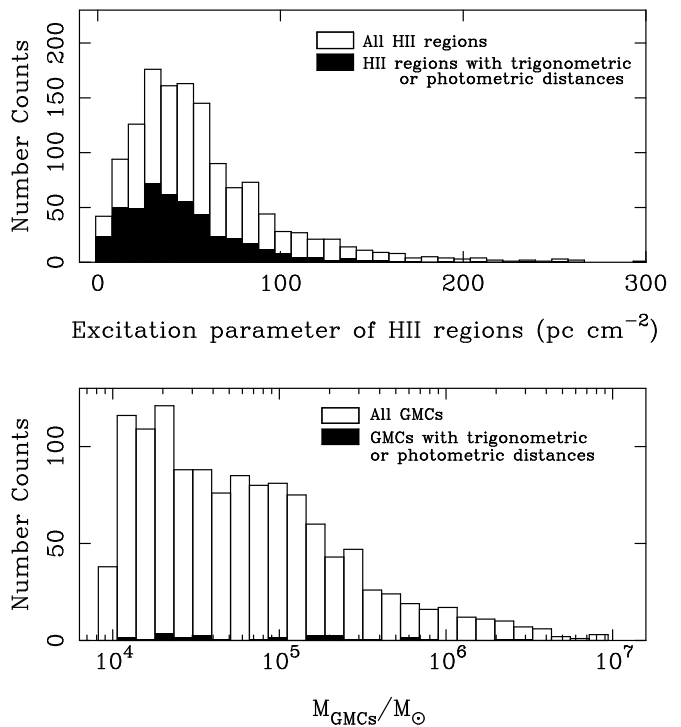}
\caption{Distributions of the excitation parameters of HII regions
  ({\it upper}) and the masses of GMCs ({\it lower}). We adopted the
  IAU standard $R_0=8.5$~kpc and $\Theta_0=220$~km~s$^{-1}$ and the
  standard solar motions together with a flat rotation curve in
  deriving the kinematic distances from velocities if no photometric
  or trigonometric distance is available.}
\label{weight}
\end{figure}

The IAU standards of solar motions with respect to the LSR are
$U_\odot=10.27$~km~s$^{-1}$, $V_\odot=15.32$~km~s$^{-1}$, and
$W_\odot=7.74$~km~s$^{-1}$ in J2000. The values of $U_\odot$ and
$W_\odot$ do not change considerably in the recent research. But the
solar motion in the direction of Galactic rotation, $V_\odot$, is
under debate. By analyzing the Hipparcos data, \citet{db98} obtained
the solar motions $U_\odot=10.00\pm0.36$~km~s$^{-1}$,
$V_\odot=5.25\pm0.62$~km~s$^{-1}$, and
$W_\odot=7.17\pm0.38$~km~s$^{-1}$. According to these parameters,
\citet{rmz+09b} analyzed the parallax measurements toward 18 massive
star forming regions, and they found a velocity difference of
$\sim$15~km~s$^{-1}$ in rotation, which may be induced by an
erroneous value of $V_\odot$. Higher values of
$V_\odot=12.24-14.6$~km~s$^{-1}$ are also suggested by some recent
studies \citep[e.g.,][]{fa09,sbd10,cab+11,rmb+14}. Based on the solar
motion parameters given by \citet[][]{sbd10}, \citet{brm+11}
reanalyzed the trigonometric parallax data of massive star forming
regions, and found the peculiar rotation velocity of
$\sim$$8\pm2$~km~s$^{-1}$.

Throughout this work, we adopt two sets of $R_0$, $\Theta_0$, and
solar motions: one is the IAU standard $R_0=8.5$~kpc and
$\Theta_0=220$~km~s$^{-1}$, and solar motions of
$U_\odot=10.27$~km~s$^{-1}$, $V_\odot=15.32$~km~s$^{-1}$, and
$W_\odot=7.74$~km~s$^{-1}$; the other is $R_0=8.3$~kpc and
$\Theta_0=239$~km~s$^{-1}$ \citep[][]{brm+11,rmb+14}, and solar
motions of $U_\odot=11.10\pm1.2$~km~s$^{-1}$,
$V_\odot=12.24\pm2.1$~km~s$^{-1}$ and $W_\odot=7.25\pm0.6$~km~s$^{-1}$
\citep{sbd10}. A flat rotation curve \citep[][]{rmz+09b,rmb+14,brm+11}
and the rotation curve of \citet[][hereafter BB93]{bb93} are adopted
in calculating the kinematic distances.

\subsection{Distances of the Galactic spiral tracers}

If the photometric or trigonometric distance is available for a
tracer, we adopt it directly. Otherwise, the kinematic distance is
estimated using the observed $V_{\rm LSR}$ and a rotation curve with
the adopted $R_0$, $\Theta_0$, and solar motions. For GMCs and 6.7 GHz
methanol masers, their kinematic distances are calculated in the same
way as that for HII regions.

Some HII regions have more than one measurement of $V_{\rm LSR}$from
the same or different emission lines (Table~\ref{tab_a1}). Here, the
mean velocity ($V_{\rm LSR}$=$\Sigma\frac{V_i}{N}$) is adopted. We
revise the mean $V_{\rm LSR}$ according to the adopted solar motions,
then calculate the kinematic distance with a flat rotation curve
\citep[][]{rmz+09b,rmb+14,brm+11} or the rotation curve of BB93. The
FORTRAN package kindly provided by \citet{rmz+09b} was modified and
used in our calculations. The sources in the range of Galactic
longitudes of $-$15$^\circ$ to 15$^\circ$ should be carefully
calculated for their kinematic distances \citep[][]{tw14} because of
the Near and Far 3 kpc Arms \citep[e.g.,][]{dt08,gcm+11,jdd+13} and
the Galactic bar(s) \citep[][]{hgm+00,cbm+09}. We first inspect their
possible associations with the Near and Far 3 kpc Arms according to
the arm parameters given by \citet[][]{dt08}, and then calculate their
distances.

Some HII regions, GMCs, and/or 6.7 GHz methanol masers are probably
associated with each other. To remove the redundancy, the associations
are identified following these criteria: (1) if an association was
identified in literature, we adopted it directly; (2) if a HII
region/maser is located within a GMC in projection considering its
measured angular size, and their velocity difference is
$\leq$~10~km~s$^{-1}$ \citep{fdt90}, we regard them as an association;
(3) if the coordinates of a HII region and a maser are closer than
2$^{\prime}$, and their velocity difference is $\leq$~10~km~s$^{-1}$
\citep{gm11}, they are regarded as association. We checked the
associations one by one, and adopted the distances of the HII regions
or GMCs in the associations preferentially for the distributions
because of the larger uncertainties of the systematic velocities derived
from the 6.7 GHz methanol maser spectra. For the associations of HII
regions and GMCs, we adopted the parameters of HII regions
(coordinates and weights) for the distributions.

\subsection{Weights for tracers of Galactic spiral structure}

\begin{figure*}
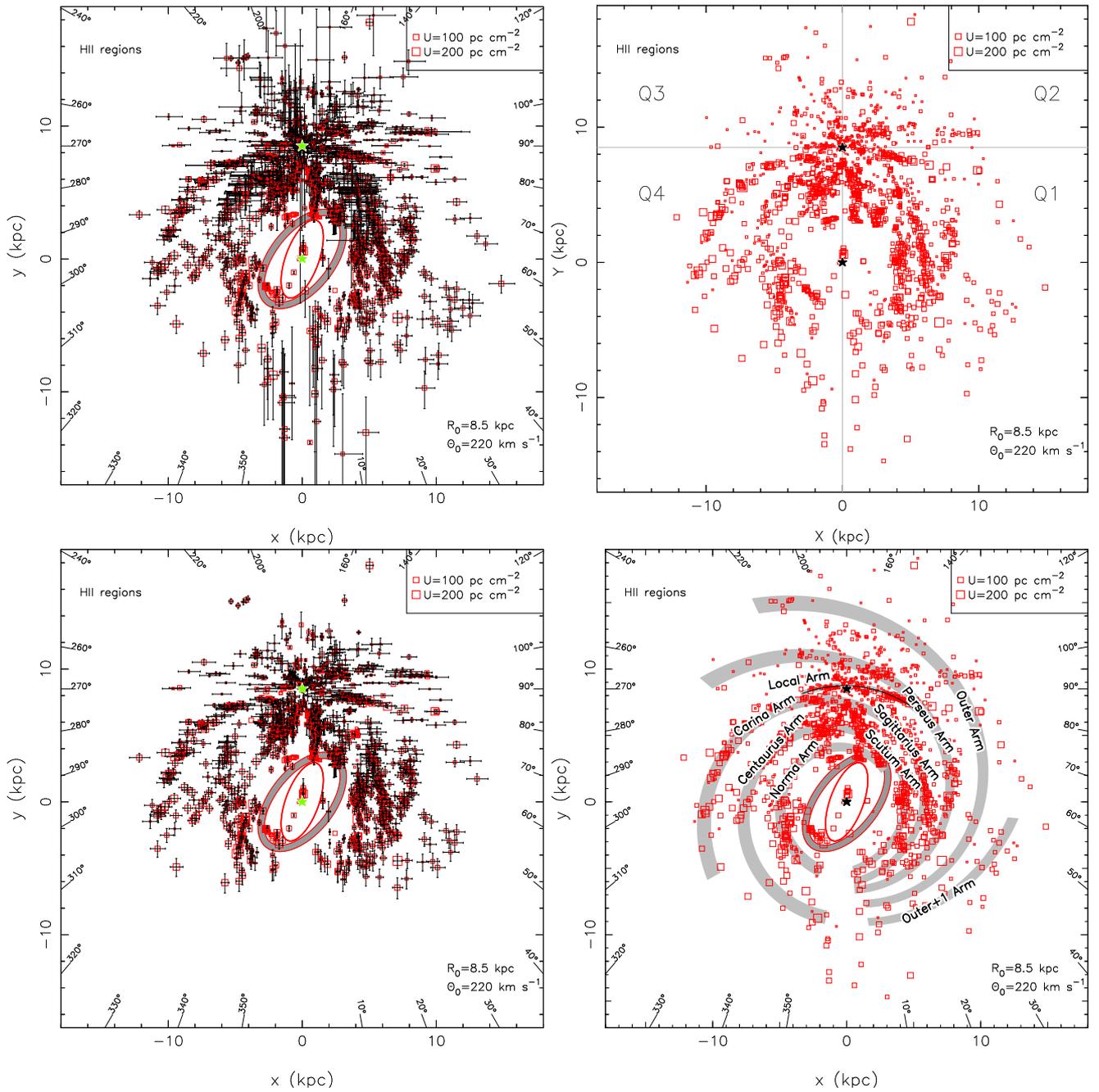

\centering\includegraphics[width=0.47\textwidth]{dis_hii_8.5_err.ps}
\centering\includegraphics[width=0.47\textwidth]{dis_hii_8.5.ps}\\
\centering\includegraphics[width=0.47\textwidth]{dis_hii_8.5_err_1kpc.ps}
\centering\includegraphics[width=0.47\textwidth]{polyarm_hii_8.5_show.ps}
\caption{{\it Upper panels}: distributions of HII regions projected
  into the Galactic plane with ({\it left}) and without ({\it right})
  position error bars. {\it Lower left}: HII region distribution
  for those only with position uncertainties better than 1~kpc. {\it
    Lower right}: HII region distribution overlaid with a spiral
  arm model (see the {\it left panel} of Fig.~\ref{model_pol}) to indicate
  the identified arm segments. The area of open squares is
  proportional to the excitation parameters (see Eq.~\ref{eq_U}).  The
  IAU standard $R_0=8.5$~kpc and $\Theta_0=220$~km~s$^{-1}$ and the
  standard solar motions together with a flat rotation curve are
  adopted in deriving the kinematic distances from velocities if no
  photometric or trigonometric distance is available. The coordinates
  originate from the GC, and the Sun is located at $x=$~0.0~kpc,
  $y=$~8.5~kpc. The open red ellipse indicates the Galactic bar
  \citep{cbm+09}, and the laurel-gray ellipse indicates the
  best-fitted Near 3 kpc Arm and Far 3 kpc Arm \citep{gcm+11}.}
\label{dis_hii}
\end{figure*}

\begin{figure*}
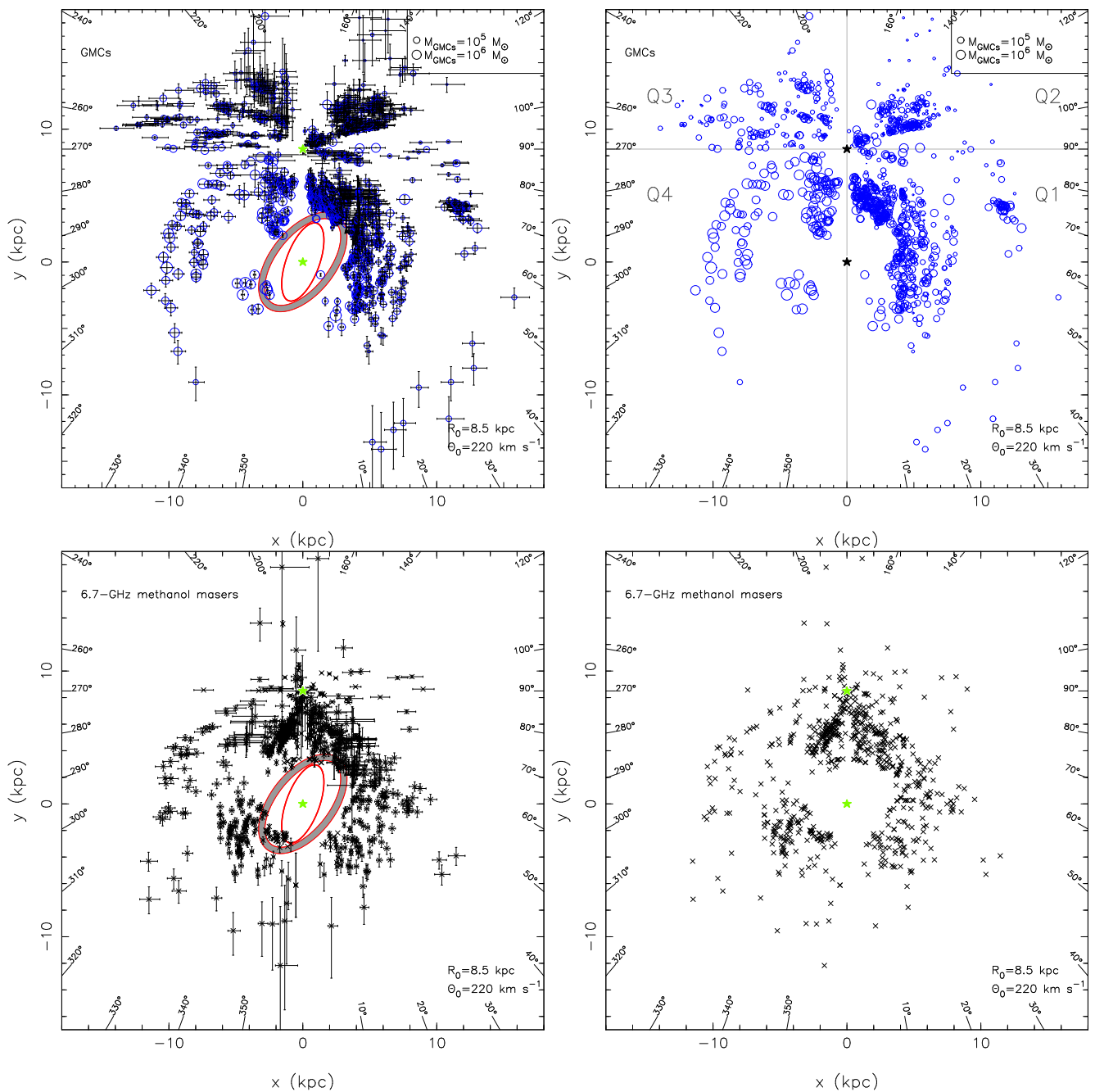

\centering\includegraphics[width=0.47\textwidth]{dis_gmc_8.5_err.ps}
\centering\includegraphics[width=0.47\textwidth]{dis_gmc_8.5.ps}\\
\centering\includegraphics[width=0.47\textwidth]{dis_maser_8.5_err.ps}
\centering\includegraphics[width=0.47\textwidth]{dis_maser_8.5.ps}\\
%
\caption{{\it Upper panels}: distributions of GMCs with ({\it left})
  and without ({\it right}) position error bars. {\it Lower panels}:
  distributions of 6.7 GHz methanol masers with ({\it left}) and
  without ({\it right}) position error bars. The symbols for GMCs and
  masers are the same as those in Fig.~\ref{lvall}. The open red
  ellipse and the laurel-gray ellipse are the same as those in
  Fig.~\ref{dis_hii}.}
\label{dis_gmc_mas}
\end{figure*}

In order to use available tracers to outline the spiral structure, an
appropriate weight factor should be used \citep{hhs09} to represent
the relative importance of a tracer for the spiral
structure. Reasonably, the brighter a HII region, or more massive a
GMC, the more important a tracer for spiral structure, and therefore a
larger weight is assigned.

For HII regions, we adopt the excitation parameter \citep{sm69} as the
weight. The excitation parameter $U$ (in pc~cm$^{-2}$) is defined as:
\begin{equation}
 U=4.5526\{\alpha(\nuup,T_e)^{-1}\nuup^{0.1}T_e^{0.35}S_{\nuup}D^{2}\}^{1/3},
\label{eq_U}
\end{equation}
here, $T_e$ is the electron temperature in K; $S_{\nu}$ is the radio
flux density in Jy, collected from literature
\citep[e.g.,][]{dwbw80,ch87,gwbe95,kc97} if available; $\nu$ is the
observed frequency in GHz; $D$ is the distance to the tracer in kpc;
and $\alpha(\nu,T_e)$ is a parameter close to 1~\citep[][]{sm69}, and
for simplicity, we adopt $\alpha(\nu,T_e)=1$. For a HII region, if
$T_e$ is given in the literature, we adopt it directly. The mean value
is used if there is more than one measurement, see
Table~\ref{tab_a1}. Otherwise, we adopt the most probable value of
$T_e=$~6500~K (see Fig.~\ref{dis_te}). The excitation parameter $U$
is calculated via Eq.~\ref{eq_U}, and the distribution of $U$ is shown
in Fig.~\ref{weight}, which ranges from 1 pc~cm$^{-2}$ to $\sim$300
pc~cm$^{-2}$ with a peak near 50 pc~cm$^{-2}$. The ranges of size or
luminosity of the Galactic HII regions are comparable to those in
other spiral galaxies \citep[e.g., M31, M51,][]{amb11}. In the catalog
of Galactic HII regions in Table~\ref{tab_a1}, ultracompact HII
regions are included, though they have much smaller sizes and radio
flux density, but they are important to indicate spiral structure in
some regions.

The weight of a HII region is defined as $W_{HII} = \frac{U}{U_0}$,
here $U_0 = 100$ pc~cm$^{-2}$. Clearly, $W_{HII}$ has a value in the
range of 0.0$-$3.0. For those HII regions without measurements of
$S_{\nu}$, a weight factor $W_{HII}=0.1$ is assigned to exploit the
distribution of entire HII region dataset.

For GMCs, we use their masses as the weight factor
\citep[see][]{hhs09}. The mass of each molecular cloud from literature
was re-scaled with the adopted distances in Table~\ref{tab_a2}, and
the distribution of $M_{GMCs}$ in the ranges of 1$\times 10^4
M_{\odot}$ to $\sim$10$^7 M_{\odot}$ is shown in the lower panel of
Fig.~\ref{weight}. The range of GMC mass in our Galaxy is comparable
to nearby spiral galaxies, e.g., M31 \citep[][]{kgf+14}. Here, we
adopt $W_{GMCs}=\log(M_{GMCs}/10^4M_{\odot})$ as the weight, and their
values are in the ranges from 0.0$-$3.0. For the methanol masers, we
simply assign the weight factor $W_{m}=0.1$ for each of them.

We also consider the weight due to the distance uncertainty of each
tracer. A weight factor $W_x=0.5/\sigma_x$ and $W_y=0.5/\sigma_y$ is
assigned to each tracer \citep{hhs09}. If the uncertainty of the
photometric or trigonometric distance is available in reference, we
adopt it directly. If the kinematic distance is used, we assume a
systematic velocity uncertainty of $\pm$~7~km s$^{-1}$, and then the
distance uncertainty is calculated via the adopted rotation curve and
the Galaxy fundamental parameters. Furthermore, if a tracer has a
distance accuracy better than 0.5 kpc in $x$ or $y$ direction, the weight
is assigned to be 1.0, i.e., $W_x$ = 1.0 if $\sigma_x< 0.5$ kpc;
and/or $W_y$ = 1.0 if $\sigma_y < 0.5$ kpc.

\begin{figure*}
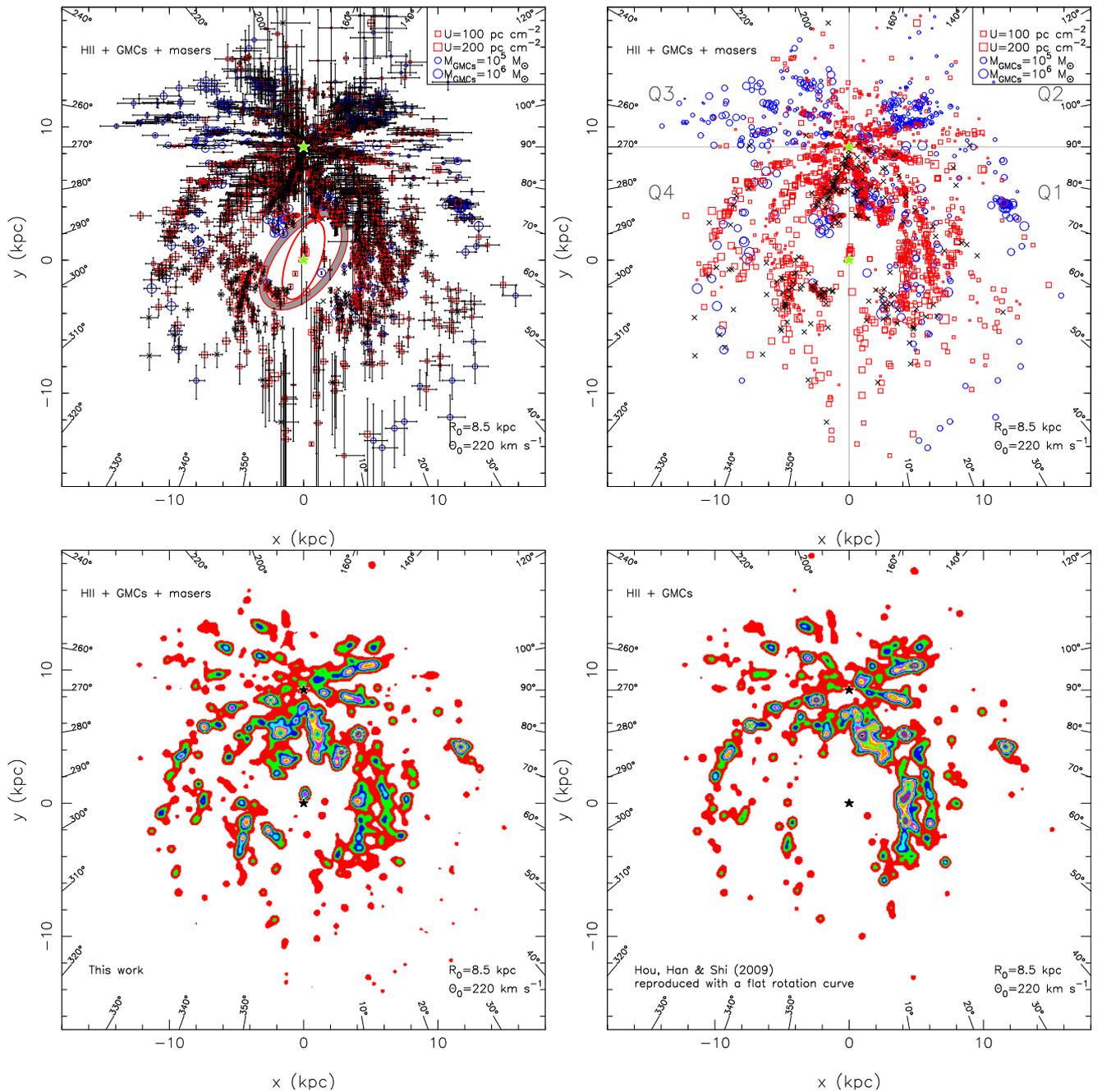

  \centering\includegraphics[width=0.47\textwidth]{dis_all_8.5_err.ps}
  \centering\includegraphics[width=0.47\textwidth]{dis_all_8.5.ps}\\
  \centering\includegraphics[width=0.47\textwidth]{Rdis_contour_8.5.ps}
  \centering\includegraphics[width=0.47\textwidth]{hhs09_8.5_flat_RC.ps} \\
  \caption{{\it Upper panels}: distributions of HII regions, GMCs, and
    6.7 GHz methanol masers with ({\it left}) and without ({\it
      right}) position error bars. The symbols for HII regions, GMCs,
    and masers are the same as those in Fig.~\ref{lvall}.  {\it
      Lower left}: color map of all tracers brightened with a Gaussian
    function with the amplitude of a weight factor, compared to the
    color map in the {\it lower right panel} which is constructed using the
    815 HII regions and 963 GMCs collected in our previous work
    \citep{hhs09}. }
\label{dis_all}
\end{figure*}

\section{The spiral structure of the Milky Way}

\subsection{Spiral structure revealed by different tracers}

\subsubsection{HII regions}

The large dataset of HII regions gives a clear presentation of the
spiral pattern, see Fig.~\ref{dis_hii}. Some remarkable features are
discussed below.

The Local Arm, where the Sun is located, is a remarkable arm segment
as indicated by a black-solid line in the lower-right panel of
Fig.~\ref{dis_hii}. It starts near the Perseus Arm ($x \sim$5 kpc,
$y \sim$7.5 kpc), and seems to have two branch-like structures near
the Sun, one to $x \sim$ $-$1 kpc and $y \sim$9 kpc close to the
Perseus Arm, above the fitted position of the Local Arm (see
Fig.~\ref{dis_hii}), the other to $x \sim$ $-$1.5 kpc and $y \sim$8.5
kpc close to the Carina Arm and overlapped with the fitted Local
Arm. \citet{xlr+13} suggested that the Local Arm is closer to the
Perseus Arm than to the Sagittarius Arm, and very likely is an arm
branch from the Perseus Arm.

In the first Galactic quadrant (Q1), there are three obvious arm
segments. From the inside out, they are the Scutum Arm, the
Sagittarius Arm and the Perseus Arm. In the outer Galaxy regions ($x$
from $\sim$7 kpc to $14$ kpc, $y$ from $\sim$ $-$10 kpc to $-$1 kpc),
there are indications for the Outer Arm and even the Outer+1 Arm. In
the inner Galaxy ($x \sim$2~kpc, $y \sim$3~kpc), the assemblies of
HII regions may be related with the near ends of the Galactic bar
and/or the 3 kpc Arms.

In the second (Q2) and third (Q3) Galactic quadrants, i.e., the
anticentral regions, a remarkable arm segment ($x$ from $-$4 kpc to 4
kpc, $y\sim$10 kpc) is the Perseus Arm. Outside the Perseus Arm,
there are the extension of the Outer Arm or even that of the Outer+1
Arm. The kinematic distance anomalies \citep{robe72} in the second and
third Galactic quadrants do not significantly influence the structure
features shown in Fig.~\ref{dis_hii} because most of the bright HII
regions have photometric or trigonometric distances
(cf. Fig.~\ref{dis_op_all}).

In the fourth Galactic quadrant (Q4), three arm segments are
obvious. From the outside in, they are the Carina Arm, the Centaurus
Arm and the Norma Arm. The assemblies of HII regions in the inner
Galaxy ($x \sim$ $-$2~kpc, $y \sim$ $-$3~kpc) maybe related to the far
ends of the Galactic bar and/or the 3 kpc Arms.

The large uncertainties of kinematic distances of HII regions may
result in some fuzzy structure features. In the lower left panel of
Fig.~\ref{dis_hii}, only the HII regions with position uncertainties
better than 1 kpc are plotted. The separation between the adjacent arm
segments is clearer, typically larger than the position uncertainties
of spiral tracers, which confirms the existence of the arm segments
discussed above.

\begin{figure}
  \centering\includegraphics[width=0.47\textwidth]{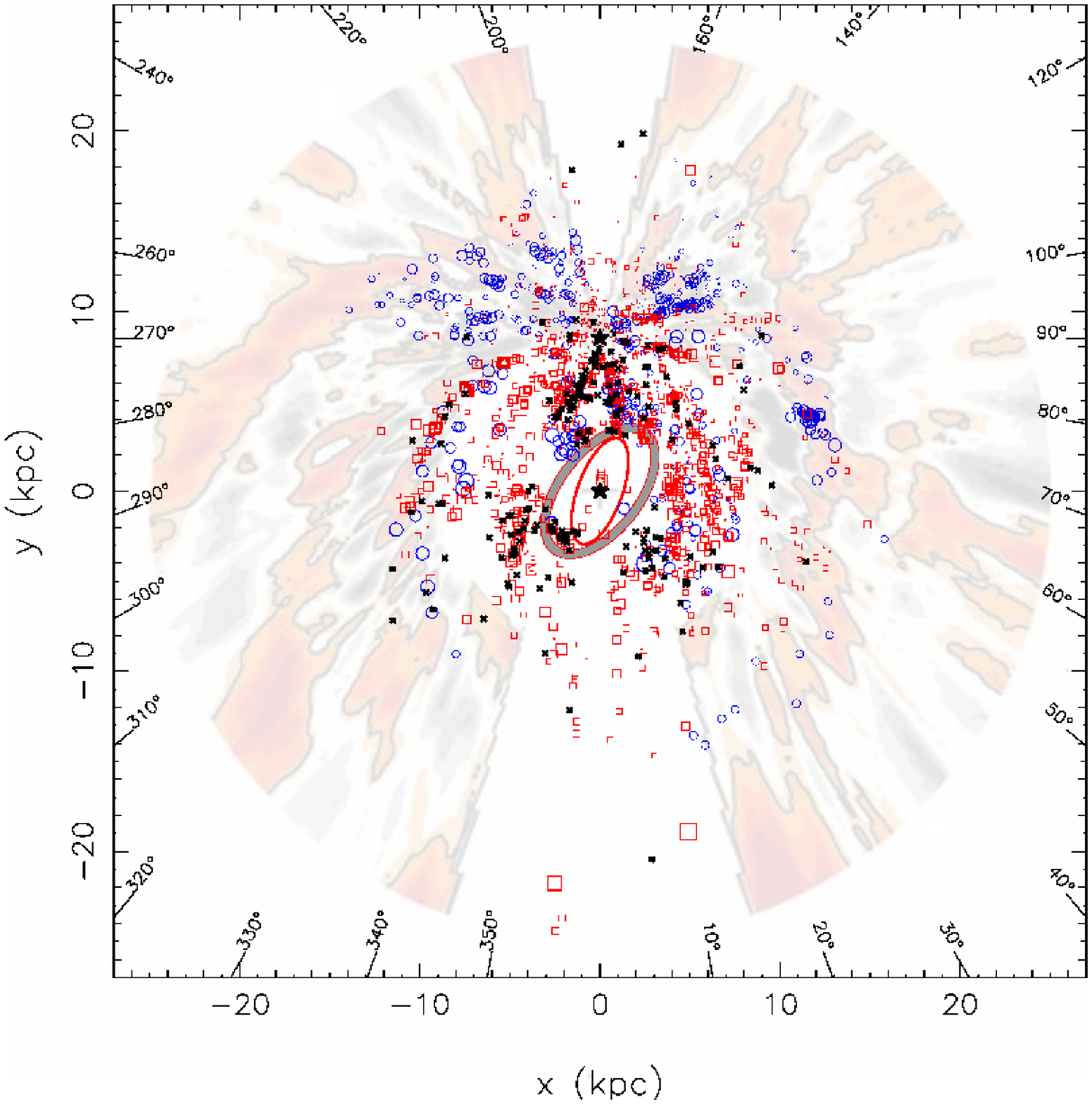}
  \caption{Spiral arm tracer distribution overlaid on the HI map
    of \citet{lbh06}. The symbols for HII regions, GMCs, and masers
    are the same as those in Fig.~\ref{lvall}.}
\label{dis_hi}
\end{figure}

\subsubsection{GMCs}

Arm features traced by GMCs are shown in
Fig.~\ref{dis_gmc_mas}. Most of the arm features recognized in the
distribution of HII regions have their counterparts in the
distribution of GMCs.

In the first Galactic quadrant, we find three arm segments, which are
the Scutum Arm, the Sagittarius Arm and the Perseus Arm from the
inside out. In the distant regions of the first quadrant
($D\gtrsim$~20~kpc, $l\sim10\degr-60\degr$), several CO emission
features were revealed by \citet{dt11}. One fully mapped CO emission
source has a mass of $\sim$5~$\times$~10$^4$ $M_\odot$. Other CO
emission features are not yet fully mapped, but are also probably from
GMCs. These GMCs may indicate the arm segment(s) beyond the Outer Arm.
Some GMCs ($x\sim$12~kpc, $y\sim$4~kpc) identified by \citet{dtb90}
may trace the Outer Arm or the Outer+1 Arm.

In the second and third Galactic quadrants many GMCs exist that were
identified by \citet{mab97}, \citet{hcs01}, and \citet{nomf05}. But
arm-like features are not obvious probably due to the anomalies of
kinematic distances \citep{robe72}, as most of their distances are
kinematic.

In the fourth quadrant, there are two obvious arm segments. From the
outside in, they are the Carina Arm and the Centaurus Arm. The
assemblies of GMCs near $x \sim$ $-$2~kpc and $y \sim$3~kpc may trace
part of the Norma Arm. The GMCs near $x \sim$ $-$3~kpc and $y \sim$
$-$2~kpc may be related to the extension of the Norma Arm and/or with
the Near 3 kpc Arm.

\subsubsection{6.7 GHz methanol masers}

The distribution of 6.7 GHz methanol masers (Fig.~\ref{dis_gmc_mas})
resembles that of HII regions or GMCs but with a much larger data
scatter. Most of their distances are kinematic, which depends on the
measured $V_{\rm LSR}$. As discussed in Sect. 2.1.3, the systematic
velocity derived from the observed maser lines may be significantly
affected by noncircular motions of the maser spots, and hence has
larger uncertainties of derived distances than those for HII regions
and GMCs.

\subsubsection{Spiral structure seen from tracers}

\begin{figure}
  \centering\includegraphics[width=0.47\textwidth]{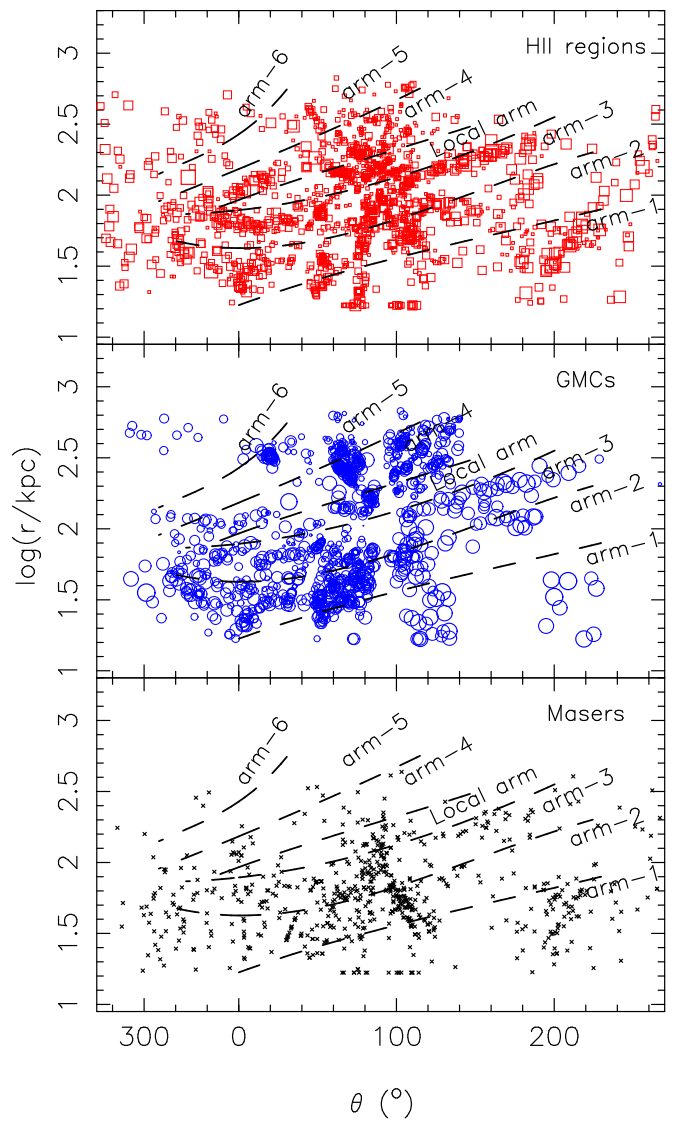}
  \caption{ Distributions of HII regions ({\it upper}), GMCs ({\it
      middle}), and 6.7 GHz methanol masers ({\it lower}) in the
    log($r$) $-$ $\theta$ diagram. The symbols for HII regions, GMCs,
    and masers are the same as those in Fig.~\ref{lvall}.  Here, $r$
    is the distance to the GC, $\theta$ starts from the positive
    $x$-axis and increases counterclockwise. The arm segments are
    identified and roughly separated by the dashed-lines. }
\label{log_seta}
\end{figure}

The distribution of three kinds of spiral tracers is shown in
Fig.~\ref{dis_all}, after removing the redundancy, as described in
Sect. 2.3.
To better indicate spiral arms, following \citet{hhs09},
we use a Gaussian function to brighten each tracer:
\begin{equation}\label{eq:LebsequeI}
        L(x,y)=\sum_i {\frac{W(i)} {2\pi
    \sigma^2}}exp(-{\frac{(x_{i}-x)^{2}+(y_{i}-y)^{2}} { 2
    \sigma^2}}).
\end{equation}
Here, $W(i)=W_{HII}$ or $W_{GMCs}$ or $W_{m}$ is the weight factor
discussed in Sect. 2.4; $x_i$ and $y_i$ are the coordinates of the
$i$th spiral tracer. A color intensity map is shown in the lower left
panel of Fig.~\ref{dis_all}, in which $\sigma$ = 0.2 is adopted as
the Gaussian width. Different values of $\sigma$ in a suitable range
(e.g., 0.05$-$0.4) yield a similar result. Several distinct arm
segments are clearly shown. Comparing to the previous arm structure in
\citet[][see the lower right panel of Fig.~\ref{dis_all}]{hhs09}, we
found that the Outer Arm, part of the Outer+1 Arm, the Local Arm, the
Centaurus Arm and the Norma Arm are significantly better outlined by
more tracers.

Comparison of the tracer distribution with the distribution of HI gas
is shown in Fig.~\ref{dis_hi}, which gives us an intuitive feeling
about the extension of the identified arm segments. In the outer
regions of the first quadrant ($x \sim$12~kpc, $y \sim$ $-$8~kpc), tens
of newly discovered HII regions \citep{abbr11} and GMCs \citep{dt11}
are consistent with an HI arm-like feature, and then connected to a
cluster of GMCs ($x \sim$12~kpc, $y \sim$4~kpc). In the third and
fourth Galactic quadrants, from the outside in, the three remarkable
HI arms correspond to the extensions of the Outer Arm ($x$ from
$\sim$ $-$13 kpc to $-$7~kpc, $y\sim$12~kpc, $l\sim200\degr-270\degr$),
the Perseus Arm ($x\sim$ $-$7~kpc, $y\sim$9~kpc, $l\sim275\degr$), and
the Carina Arm ($x\sim$ $-$10 kpc). In particular, the Carina Arm traced
by HII regions and GMCs is well matched with distribution of HI gas.

In summary, the distribution of HII regions and GMCs do show some
obvious arm segments. Particularly, these arm segments are able to
match the HI arms in the outer Galaxy. However, the connections of arm
segments in different Galactic quadrants are still not intuitive.

\subsection{Fitting models to tracer distributions }

We can roughly identify the arm segments in the log($r$)$-$$\theta$
diagrams as separated by the dashed lines in
Fig.~\ref{log_seta}. From the bottom up, they are the Norma Arm
(arm-1), the Scutum-Centaurus Arm (arm-2), the Sagittarius-Carina Arm
(arm-3), the Local Arm, the Perseus Arm (arm-4), the Outer Arm (arm-5)
and the Outer+1 Arm (arm-6). Near the direction of $\theta \sim$90$^\circ$, 
the spiral tracers are mixed, the Scutum-Centaurus Arm and
the Sagittarius-Carina Arm can barely be distinguished. Near the
direction of $\theta \sim$320$^\circ$, where the data points are also
mixed, the different arm segments cannot be separated with high
confidence. In addition, there is a lack of tracers at large $r$, the
Outer Arm and the Outer+1 Arm can only be identified marginally. The
Local Arm ($\theta \sim$$50^\circ-110^\circ$, log($r$) $\sim$2.1) is
obviously short and has a distinctive pitch angle as recognized in the
distribution of HII regions. To explore the number and position of
spiral arms in our Galaxy, fitting the tracer data with models is
desired.

Many models of spiral structure have been proposed
\citep[e.g.,][]{val08,val13}. The paradigmatic model is the four arm
segments first recommended by GG76, which is supported by observations
later~\citep[e.g.,][]{dwbw80,ch87,eg99,ds01,rus03,lbh06,bb14}, though
other spiral patterns have been proposed to describe the structure of
the Galaxy \citep[see, e.g.,][]{al97,lmd01,lra+11,peb08,hhs09}. Here,
we first fit the tracer data with the conventional logarithmic
spirals, and then present a model with polynomial-logarithmic spiral
arms to connect the tracers.

\begin{figure*}
\centering\includegraphics[width=0.47\textwidth]{3arm_hii_8.5.ps}
\centering\includegraphics[width=0.47\textwidth]{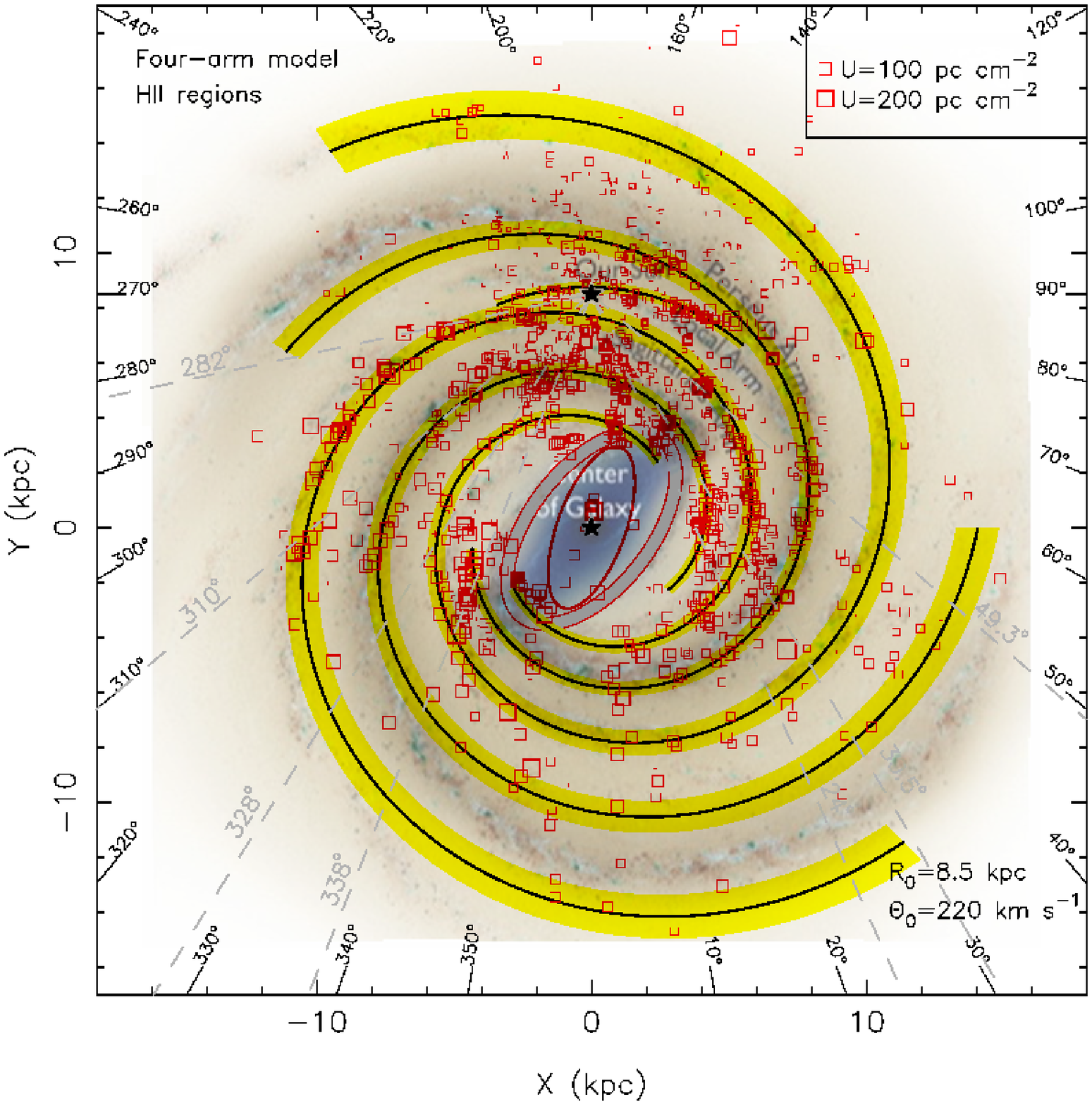}\\
\centering\includegraphics[width=0.47\textwidth]{3arm_all_8.5.ps}
\centering\includegraphics[width=0.47\textwidth]{4arm_all_8.5.ps}\\
\caption{{\it Upper panels}: best-fitted three-arm model ({\it
    left}) and four-arm model ({\it right}) for the distribution of
  HII regions. {\it Lower panels}: best-fitted three-arm model
  ({\it left}) and four-arm model ({\it right}) for the distribution
  of three kinds of spiral tracers. The symbols are the same as those
  in Fig.~\ref{lvall}. The open red ellipse indicates the Galactic
  bar \citep{cbm+09}, and the laurel-gray ellipse shows the
  best-fitted Near 3 kpc Arm and Far 3 kpc Arm \citep{gcm+11}. The
  dashed-lines indicate the observed tangential directions (see
  Table~\ref{tan_obs}).}
\label{model_log}
\end{figure*}

\setlength{\tabcolsep}{0.8mm}
\begin{table*}[!t]
  \caption{Arm parameters of the best-fitted models of logarithmic
    spirals (see Eq.~\ref{eq_log}).}
\begin{center}
\begin{tabular}{lcccccccccccccccc}
  \hline
  \hline
  Models$\setminus$Arm parameters & $R_{1}$&$\theta_{1}$& $\psiup_{1}$ & $R_{2}$ & $\theta_{2}$ &
  $\psiup_{2}$ & $R_{3}$ & $\theta_{3}$ & $\psiup_{3}$ & $R_{4}$ &
  $\theta_{4}$& $\psiup_{4}$ & $R_{5}$  & $\theta_{5}$ & $\psiup_{5}$& $Z$  \\
  & (kpc)&$(^{\circ})$& $(^{\circ})$ & (kpc) & $(^{\circ})$ &
  $(^{\circ})$ & (kpc) & $(^{\circ})$ & $(^{\circ})$  &
  (kpc)& $(^{\circ})$& $(^{\circ})$& (kpc) &$(^{\circ})$ & $(^{\circ})$&\\
  \hline
  \hline
  \multicolumn{14}{l}{Fitting of models to HII regions, $R_0=8.5$~kpc, $\Theta_0=220$~km~s$^{-1}$.} \\
  3-arm model& {3.42} & {44.4} & { 8.85} & {3.72} & {192.3} &  {9.29} & {3.01} & {219.4} &  {8.62}& & & & {8.61}& {51.3}& {0.50} &{0.17}  \\
  4-arm model& {3.40} & {44.9} & {11.70} & {4.41} & {190.1} & {10.82} & {3.58} & {217.6} & {10.68} & {3.60} &
  {320.9} & {11.06} & {8.64} & {50.4} & {0.99} & {0.16} \\
  \hline
  \multicolumn{14}{l}{Fitting of models to all three kinds of spiral tracers (HII regions, GMCs, and 6.7 GHz methanol masers).} \\
  3-arm model& {3.40} & {45.8} &  {9.12} & {3.78} & {194.7} &  {9.29} & {3.47} & {230.4} & {7.61}& & & & {8.59}& {54.7}& {0.62} &{0.20}  \\
  4-arm model& {3.23} & {41.2} & {10.64} & {4.27} & {189.0} & {11.15} & {3.56} & {217.9} & {10.63} & {3.81} &
  {309.8} & {8.85} & {8.64} & {53.8} & {0.59} & {0.17} \\
  \hline\hline
  \multicolumn{14}{l}{Fitting of models to HII regions, $R_0=8.3$~kpc, $\Theta_0=239$~km~s$^{-1}$.} \\
  3-arm model& {3.19} & {36.8} & {9.00} & {3.27} & {191.9} & {10.36} & {3.13} & {230.7} & {8.16}& & & & {8.20}& {51.1}& {2.63} & {0.16}  \\
  4-arm model& { 3.35} & {44.4} & {11.43} & {4.61} & {192.4} & {9.84} & {3.56} & {218.6} & {10.38} & {3.67} &
  {330.3} & {10.54} & {8.21} & {55.1} & {2.77}  & {0.16} \\
  \hline
  \multicolumn{14}{l}{Fitting of models to all three kinds of spiral tracers.} \\
  3-arm model& {3.22} & {44.1} & {9.25} & {3.43} & {184.5} & {9.50} & {3.10} & {210.3} & {7.80} & & & & {8.17} & {47.8} & {2.68} &{0.18}  \\
  4-arm model& {3.27} & {38.5} & {9.87} & {4.29} & {189.0} & {10.51} & {3.58} & {215.2} & {10.01} & {3.98} &
  {320.1} & {8.14} & {8.16} & {50.6} & {2.71}  & {0.16} \\
  \hline\hline
\end{tabular}
\end{center}
\label{34para}
\tablefoot{{\bf Notes.} For the $i$th spiral arm, $R_{i}$
    is the initial radius, $\theta_{i}$ is the start azimuth angle,
    and $\psiup_{i}$ is the pitch angle. The parameters of the Local
    Arm are denoted with subscript 5. The fitting factor $Z$ (see
    Eq.~\ref{eq1}) for each model is shown in the last column.}
\end{table*}

\begin{table*}
  \caption{Tangential directions of spiral arms derived from
    observations (part of the references are taken from the collection
    by Englmaier \& Gerhard 1999 as shown in their Table 1). }
\begin{center}
\begin{tabular}{lccccccc}
  \hline
  \hline
 Tracers and References   & Near 3 kpc & Scutum & Sagittarius & Carina & Centaurus & Norma & Far 3 kpc \\
  & $(^{\circ})$  & $(^{\circ})$ & $(^{\circ})$ & $(^{\circ})$ & $(^{\circ})$& $(^{\circ})$ & $(^{\circ})$   \\
  \hline
HI: wea70, bs70, hen77  &     &      & 50   &     & 310  & 328 &        \\
integrated $^{12}$CO: ccdt80, gcb+87  & 24  & 30.5 & 49.5 &     & 310  & 330 &       \\
$^{12}$CO clouds: dect86  & 25  & 32   & 51   &     &      &     &        \\
warm CO clouds: ssr85  & 25  & 30   & 49   &     &      &     &       \\
HII regions: loc89, dwbw80  & 24  & 30   & 47   &     & 305  & 332 &       \\
$^{26}$Al: cgdh96  &     &      & 46   &     & 310  & 325 &       \\
Radio 408 MHz: bkb85  &     &      & 48   &     &310,302& 328 & 339  \\
2.4$\mu$m: hmm+81  &     &      &      &     &      & 332 & 339    \\
60$\mu$m: bdt90  &     &      &      &     & 313  & 329 & 340      \\
$^{12}$CO: bro92  &25   & 33   & 55   & 282 & 309  & 328 & 337  \\
$^{12}$CO: bro08  &     &      &      &     & 308  & 328 & 336   \\
$^{12}$CO: ban80, cd76  &23.5 &      &      &     &      &     &      \\
$^{12}$CO: dt08  &23   &      &      &     &      &     & 337  \\
\hline
  Median & 24  & 30.5 & 49.3 & 282 & 310 & 328 & 338  \\
  \hline\hline
\end{tabular}
\end{center}
\label{tan_obs}
\tablefoot{{\bf Notes.} The median values are adopted in this work if
  there is more than one measurement. bs70: \citet{bs70}; bkb85:
  \citet{bkb85}; bdt90: \citet{bdt90}; bro92: \citet{bro92}; bro08:
  \citet{bron08}; ban80: \citet{bani80}; ccdt80: \citet{ccdt80};
  cgdh96: \citet{cgdh96}; cd76: \citet{cd76}; dect86: \citet{dect86};
  ssr85: \citet{ssr85}; dt08: \citet{dt08}; dwbw80: \citet{dwbw80};
  hen77: \citet{hend77}; hmm+81: \citet{hmm+81}; gcb+87:
  \citet{gcb+87}; loc89: \citet{lock89}; wea70: \citet{weav70}.}
\end{table*}

\subsubsection{The logarithmic spiral arm models}
\label{log_mod}

Spiral arms of galaxies have been conventionally approximated by
logarithmic spirals~\citep[e.g.,][and references therein]{rus03}. For
the Milky Way, the two-arm logarithmic spirals can neither fit the
distribution of massive star forming regions/GMCs nor match the
observed tangential directions of spiral arms \citep{rus03,hhs09}. The
models of three- and four-arm spirals can fit data equally well
\citep{rus03,hhs09}. In this work, we focus on the spiral structure
traced by massive star forming regions and GMCs. As shown in some
galaxies, the spiral structure traced by the stellar component may be
different from that traced by massive star forming regions or gas
\citep[e.g.,][]{gp98}. For the Milky Way, the stellar component is
commonly accepted to be dominated by a two-arm spiral
pattern~\citep[][]{ds01,bcb05,cbm+09,fa12}.

In polar coordinates ($r,\theta$), the $i$th arm can be given as
logarithmic form:
\begin{equation}
\ln\frac{r}{R_{i}}=(\theta-\theta_{i})\tan\psiup_{i},
\label{eq_log}
\end{equation}
where $r$, $\theta$ are the polar coordinates centered at the GC,
$\theta$ starts at the positive $x$-axis and increases counterclockwise;
$\theta_{i}$, $R_{i}$, and $\psiup_{i}$ are the start azimuth angle,
the initial radius, and the pitch angle for the $i$th spiral arm,
respectively. To search for the optimized values of each parameter, we
minimize the factor \citep[see][]{rus03,hhs09}:
\begin{equation}\label{eq1}
  Z=\frac{1}{\sum{W_i}}\sum
  W_{i}\sqrt{{(x_{i}-x_{t})^{2}} {W_{x_i}^2} +
    {(y_{i}-y_{t})^{2}} {W_{y_i}^2}},
\end{equation}
here, $W_{i}$ is the weight discussed in Sect. 2.4; $x_{i}$ and
$y_{i}$ are the Cartesian coordinates of a spiral tracer, their
uncertainties are represented by $W_{x_i}$ and $W_{y_i}$,
respectively; $x_{t}$ and $y_{t}$ are the coordinates of the nearest
point from the fitted spiral arms to the tracer. The Minuit package,
with the $simplex$ and $migrad$ search routines (Nelder \& Mead 1965),
is adopted to minimize the factor $Z$.

The model is first fitted to only the HII regions, and then to the all
three kinds of spiral tracers (i.e., HII regions, GMCs, and 6.7 GHz
methanol masers). We realized that the distances of many HII regions
have been determined more credibly by photometric or trigonometric
method, especially for those in the second and third Galactic
quadrants where the kinematic anomalies dominate. The known HII
regions are widely distributed in a large region of the disk (see
Fig.~\ref{dis_hii} and \ref{dis_gmc_mas}).

\begin{table*}
  \caption{Tangential directions of spiral arms derived from the
    best-fitted models. }
\begin{center}
\begin{tabular}{lcccccccc}
  \hline
  \hline
  Models$\setminus$Spiral arms & Near 3 kpc & Scutum & Sagittarius & Carina & Centaurus & Norma & Far 3 kpc &  $\sqrt{\Sigma_{i}(\phi_{i}^{model}-\phi_{i}^{obs})^{2}}$/N\\
      & 24$^{\circ}$  & 30.5$^{\circ}$ & 49.3$^{\circ}$ & 282$^{\circ}$ & 310$^{\circ}$ & 328$^{\circ}$ & 338$^{\circ}$  \\
  \hline
  \hline
  \multicolumn{8}{l}{Best-fitted models to HII regions, $R_0=8.5$~kpc, $\Theta_0=220$~km~s$^{-1}$.}  &  \\
  3-arm model & {22.6} [1.4] & {34.7} [4.2]& {57.5} [8.2]& {285.0} [3.0]& {312.5} [2.5]& {327.4} [0.6]&   & {1.7}\\
  4-arm model & {22.1} [1.9] & {33.6} [3.1]& {55.3} [6.0]& {284.7} [2.7]& {309.8} [0.2]& {324.2} [3.8]&   & {1.4} \\
  PL model    & {24.3} [0.3] & {33.3} [2.8]& {51.1} [1.8]& {286.1} [4.1]& {310.2} [0.2]& {326.8} [1.2]&  & {0.9}\\
  \hline
  \multicolumn{8}{l}{Best-fitted models to all three kinds of spiral tracers.} \\
  3-arm model &{22.3} [1.7]  & {37.5} [7.0]& {58.7} [9.4]& {283.7} [1.7]& {310.9} [0.9]& {327.4} [0.6]&   & {2.0}\\
  4-arm model &{21.2} [2.8]  & {34.9} [4.4]& {54.1} [4.8]& {286.1} [4.1]& {311.9} [1.9]& {326.8} [1.2]&   & {1.4} \\
  PL model    &{24.9} [0.9]  & {33.6} [3.1]& {50.8} [1.5]& {286.5} [4.5]& {310.1} [0.1]& {326.6} [1.4]&  & {1.0}\\
  \hline\hline
  \multicolumn{8}{l}{Best-fitted models to HII regions, $R_0=8.3$~kpc, $\Theta_0=239$~km~s$^{-1}$.}  \\
  3-arm model &{21.9} [2.1]  & {35.1} [4.6]& {54.6} [5.3]& {286.1} [4.1]& {312.8} [2.8]& {328.0} [0.0]&   & {1.5}\\
  4-arm model &{22.4} [1.6]  & {33.5} [3.0]& {54.7} [5.4]& {285.9} [3.9]& {310.8} [0.8]& {324.1} [3.9]&   & {1.4} \\
  PL model    &{23.8} [0.2]  & {32.5} [2.0]& {49.7} [0.4]& {288.9} [6.9]& {310.2} [0.2]& {327.3} [0.7]&  &{1.2} \\
  \hline
  \multicolumn{8}{l}{Best-fitted models to all three kinds of spiral tracers.} \\
  3-arm model &{21.6} [2.4]  & {36.0} [5.5]& {55.3} [6.0]& {287.0} [5.0]& {312.3} [2.3]& {328.1} [0.1]&   & {1.7}\\
  4-arm model &{22.4} [1.6]  & {35.8} [5.3]& {53.9} [4.6]& {287.6} [5.6]& {311.9} [1.9]& {326.3} [1.7]&  & {1.6} \\
  PL model    &{22.9} [1.1]  & {32.7} [2.2]& {51.7} [2.4]& {288.7} [6.7]& {310.6} [0.6]& {327.9} [0.1]&  & {1.3} \\
  \hline\hline
\end{tabular}
\end{center}
\label{tan_mod}
\tablefoot{{\bf Notes.} The deviations between the modeled and
  observed tangential directions ($|\phi_{i}^{model}-\phi_{i}^{obs}|$,
  see Table~\ref{tan_obs} for $\phi_{i}^{obs}$) are given in
  brackets.}
\end{table*}

\begin{figure*}
  \centering\includegraphics[width=0.47\textwidth]{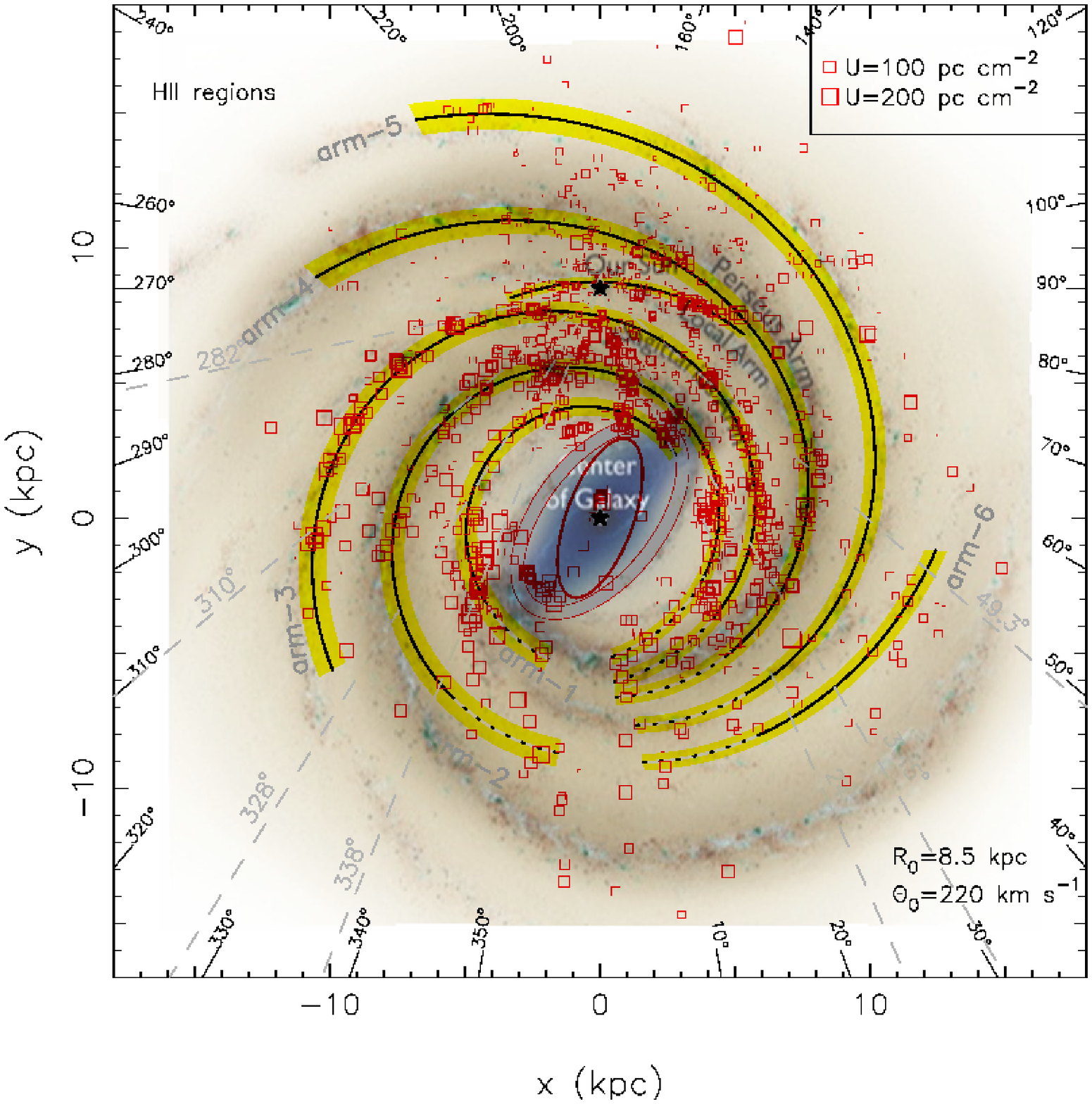}
  \centering\includegraphics[width=0.47\textwidth]{polyarm_all_8.5.ps}\\
  \caption{ Best-fitted models of polynomial-logarithmic (PL) spirals
    to only the HII regions ({\it left}) and all three kinds of spiral
    tracers ({\it right}). The symbols are the same as those in
    Fig.~\ref{lvall}. The dashed-lines indicate the observed
    tangential directions (see Table~\ref{tan_obs}). In the far end of
    the GC direction, the spiral tracers generally have large position
    error bars (see e.g., Fig.~\ref{dis_hii}), resulting in the large
    uncertainties of the fitted positions of arm segments which are
    indicated as dashed-lines for Galaxy longitude range of 340$^\circ
    \leq$ $l$ $\leq$20$^\circ$.}
\label{model_pol}
\end{figure*}

The best-fitted models to HII regions are displayed in the upper
panels of Fig.~\ref{model_log}, and the corresponding parameters of
spiral arms are listed in Table~\ref{34para}. Both the three-arm model
and the four-arm model can connect most spiral tracers. The fitting
factor $Z$ = 0.17 for the best-fitted three-arm model is slightly
larger than that for the four-arm model ($Z$ = 0.16). However, the
connections of the known arm segments differ in the two models. In the
four-arm model, the Norma Arm starts from the near end of the Galactic
bar, then extends to the Outer Arm. While in the three-arm model, the
Norma Arm is connected to the Perseus Arm, which is fitted as an
individual spiral arm in the four-arm model. The Scutum Arm in the
four-arm model starts from a cluster of HII regions ($x\sim$4 kpc,
$y\sim$0 kpc) connected to the Centaurus Arm, then extends to be the
Outer+1 Arm. In the three-arm model, however, the Scutum Arm starts
from the far end of the Galactic bar, then connects to the Centaurus
Arm, and extends to the Outer Arm. The Sagittarius-Carina Arm is well
fitted in both of the two models, but their extensions are
different. The Local Arm is fitted additionally in both of the two
models as a short arm segment, which starts near the Perseus Arm, and
then extends to the fourth quadrant until it approaches the Carina Arm.

The tangential directions of spiral arms obtained by the best-fitted
four-arm model are slightly more consistent with observations than
those from the three-arm model, as indicated by
$\sqrt{\Sigma_i(\phi_{i}^{model}-\phi_{i}^{obs})^{2}}$/N in
Table~\ref{tan_mod}, though the deviations can be as large as
4$^{\circ}$$-$6$^{\circ}$ for the Norma Arm and the Sagittarius
Arm. Most grand-design spiral galaxies approximate two-fold rotational
symmetry~\citep[e.g,][]{kbc+92,rkb98,bt08}. For our Galaxy, a two-fold
symmetry of the spiral structure has been proposed \citep{dam13},
which is based on analyzing the observational data of stellar
component~\citep[e.g.,][]{bcb05} and the discovery of the Far 3 kpc
Arm \citep[][]{dt08} and distant GMCs and HII regions
\citep[][]{dt11,abbr11}. The two-fold symmetry is more satisfied by
the best-fitted four-arm model than the three-arm model.

The best-fitted models to all three kinds of tracers are displayed in
the lower two panels of Fig.~\ref{model_log}. The modeled Norma Arm,
Scutum-Centaurus Arm, Sagittarius-Carina Arm, and Perseus Arm are
consistent with the models from the fitting to HII regions. The
disagreements are found for the outer Galaxy regions, in particular in
the second and third quadrants, where many GMCs exist. Their
distributions are somewhat messy, because of large uncertainties in
the kinematic distances. Exact distances should be measured
\citep[e.g., W3OH in the Perseus Arm,][]{xrzm06}, otherwise the
structure features traced by many GMCs in the second and third
quadrants are questionable. In contrast, most of the bright HII
regions in the second and third quadrants have photometric or
trigonometric distances (see Fig.~\ref{dis_op_all}). The arm features
traced by HII regions are reliable, which motivates us to recommend
the best-fitted four-arm model shown in the upper right panel of
Fig.~\ref{model_log}.

\begin{table}
  \caption{Arm parameters for the best-fitted PL spiral arm models
    (see Eq.~\ref{eq_pol}).}
\begin{center}
\begin{tabular}{lcccccc}
  \hline
  \hline
  $i$th arm & $a_i$ & $b_i$ & $c_i$ & $d_i$ & $\theta_{start}$ & $\theta_{end}$ \\
             &      &       &       &       & ($^\circ$) & ($^\circ$) \\
  \hline
  \hline
  \multicolumn{6}{l}{HII regions, $R_0=8.5$~kpc, $\Theta_0=220$~km~s$^{-1}$, $Z$ = {0.16}.} \\
  arm-1 & {1.1923} & { 0.1499} & {-0.007056} & {0.0}       & 40 & 250    \\
  arm-2 & {7.4144} & {-2.4119}  & {0.3105}  & {-0.01223} & 275  & 620   \\
  arm-3 &{ 6.8191} & {-2.1630}  & {0.2888}  & {-0.01163} & 275  & 570   \\
  arm-4 & {2.6019} & {-0.3315}& {  0.03829} & {0.0}     & 275   & 500   \\
  arm-5 & {1.7840} & {-0.04095}  & {0.01956} & {0.0}     & 280  & 475   \\
  arm-6 & {3.1816} & {-0.5408}   & {0.07024} & {0.0}     & 280  & 355   \\
  & $R_i$ (kpc)   & $\theta_i$ ($^{\circ}$)   & $\psiup_i$ ($^{\circ}$)   &    &  & \\
  Local Arm  & {8.64} & {52.0} & {1.00} &    &  & \\
  \hline
  \multicolumn{6}{l}{All three kinds of spiral tracers, $Z$ = {0.18}.} \\
  arm-1 & {1.2170} & { 0.1442} & {-0.007552} &  {0.0}     & 40  & 250    \\
  arm-2 & {7.4413} & {-2.4138}  & {0.3103}  & {-0.01222} & 275  & 620   \\
  arm-3 &{ 6.8185} & {-2.1632}  & {0.2887}  & {-0.01162} & 280  & 625   \\
  arm-4 & {1.8419} & {-0.09367}& {0.02001} &  {0.0}     & 280   & 500   \\
  arm-5 & {1.7020} & {-0.01485}  & {0.01522} &  {0.0}     & 280 & 500   \\
  arm-6 & {2.0497} & {-0.05168}   & {0.01807} & {0.0}     & 280 & 405   \\
  & $R_i$ (kpc)   & $\theta_i$ ($^{\circ}$)   & $\psiup_i$ ($^{\circ}$)   &    &  & \\
  Local Arm  & {8.60} & {51.9} & {0.76} &    &  & \\
  \hline
  \hline
  \multicolumn{6}{l}{HII regions, $R_0=8.3$~kpc, $\Theta_0=239$~km~s$^{-1}$, $Z$ = {0.16}.} \\
  arm-1 & {1.1668} & { 0.1198} & { 0.002557} &  {0.0}     & 40  & 250    \\
  arm-2 & {5.8002} & {-1.8188}  & {0.2352}  & {-0.008999} & 275  & 620   \\
  arm-3 &{ 4.2300} & {-1.1505}  & {0.1561}  & {-0.005898} & 275  & 570   \\
  arm-4 & {0.9744} & { 0.1405}& {0.003995} &  {0.0}     & 280   & 500   \\
  arm-5 & {0.9887} & { 0.1714}  & {0.004358} &  {0.0}     & 280 & 475   \\
  arm-6 & {3.3846} & {-0.6554}   & {0.08170} & {0.0}     & 280 & 355   \\
  & $R_i$ (kpc)   & $\theta_i$ ($^{\circ}$)   & $\psiup_i$ ($^{\circ}$)   &    &  & \\
  Local Arm  & {8.10} & {52.2} & {2.36} &    &  & \\
  \hline
  \multicolumn{6}{l}{All three kinds of spiral tracers, $Z$ = {0.17}.} \\
  arm-1 & {1.1320} & { 0.1233} & {0.003488} &  {0.0}     & 40  & 250 \\
  arm-2 & {5.8243} & {-1.8196}  & {0.2350}  & {-0.009011} & 275 & 620 \\
  arm-3 &{ 4.2767} & {-1.1507}  & {0.1570}  & {-0.006078} & 275 & 575 \\
  arm-4 & {1.1280} & { 0.1282}& {0.002617} &  {0.0}     & 280 & 500  \\
  arm-5 & {1.7978} & {-0.04738}  & {0.01684} &  {0.0}     & 280 & 500 \\
  arm-6 & {2.4225} & {-0.1636}  & {0.02494} &  {0.0}     & 280 & 405 \\
  & $R_i$ (kpc)   & $\theta_i$ ($^{\circ}$)   & $\psiup_i$ ($^{\circ}$)   &    &  & \\
  Local Arm  & {8.17} & {57.8} & {2.84} &    &  & \\
  \hline\hline
\end{tabular}
\end{center}
\label{5arm}
\tablefoot{{\bf Notes.}  The fitting is made to HII regions and all
  three kinds of spiral tracers, respectively.}
\end{table}

\begin{figure}[!t]
\centering\includegraphics[width=0.47\textwidth]{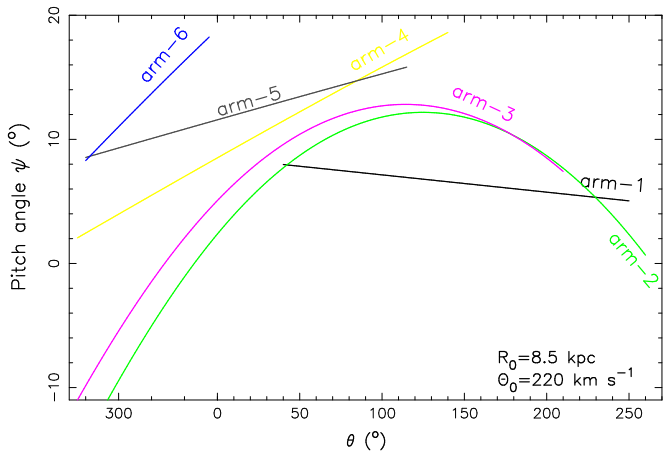}
\caption{Pitch angle ($\psiup_{i}$) variation of arm segments in
  the best-fitted polynomial-logarithmic model ({\it left panel} of
  Fig.~\ref{model_pol}).}
\label{pitch_a}
\end{figure}

\subsubsection{The polynomial-logarithmic spiral arm model}

A single value of the pitch angle cannot well describe the spiral arms
for many galaxies \citep[e.g.,][]{sj98,sr13}, and the variation of
pitch angle can sometimes exceed 20 percent \citep[][]{sr13}. To fit
both the data distribution and also the observed tangential
directions, we proposed a polynomial-logarithmic (PL) spiral arm model
\citep{hhs09} to connect the identified arm segments.

In polar coordinates ($r$, $\theta$), a PL spiral arm is expressed as:
\begin{equation}
\ln r = a_i + b_i \theta+c_i\theta^2+d_i\theta^3.
\label{eq_pol}
\end{equation}
To search the best-fitted models, the Minuit package is adopted to
minimize the fitting factor $Z$ (see Eq~\ref{eq1}). We first fit the
six identified arm segments (see the log($r$)$-$$\theta$ diagrams in
Fig.~\ref{log_seta}) individually by a PL spiral to obtain the
initial values of arm parameters. Then, the six arm segments and the
Local Arm are fitted together to derive the best model. As a purely
logarithmic spiral has shown to be good enough to fit the Local Arm
(Fig.~\ref{model_log}), we adopt it here as well. The model is also
fitted to the HII regions and to all three kinds of spiral tracers.

The best-fitted model to HII regions is shown in the left panel of
Fig.~\ref{model_pol}, the corresponding arm parameters are listed in
Table~\ref{5arm}. This model has advantages in connecting most known
spiral tracers. From the inside out, the Norma Arm (arm-1), the
Scutum-Centaurus Arm (arm-2), the Sagittarius-Carina Arm (arm-3), the
Perseus Arm (arm-4), and the Local Arm both are well fitted. The Outer
Arm (arm-5) and the Outer+1 Arm (arm-6) are also delineated, although
less spiral tracers are found for the outer Galaxy. The modeled
tangential directions for the PL arm models are in better agreement
with observations than those of the best-fitted four-arm model, 0.9
cf. 1.4, see Table~\ref{tan_mod}.

\begin{figure*}
\centering\includegraphics[width=0.47\textwidth]{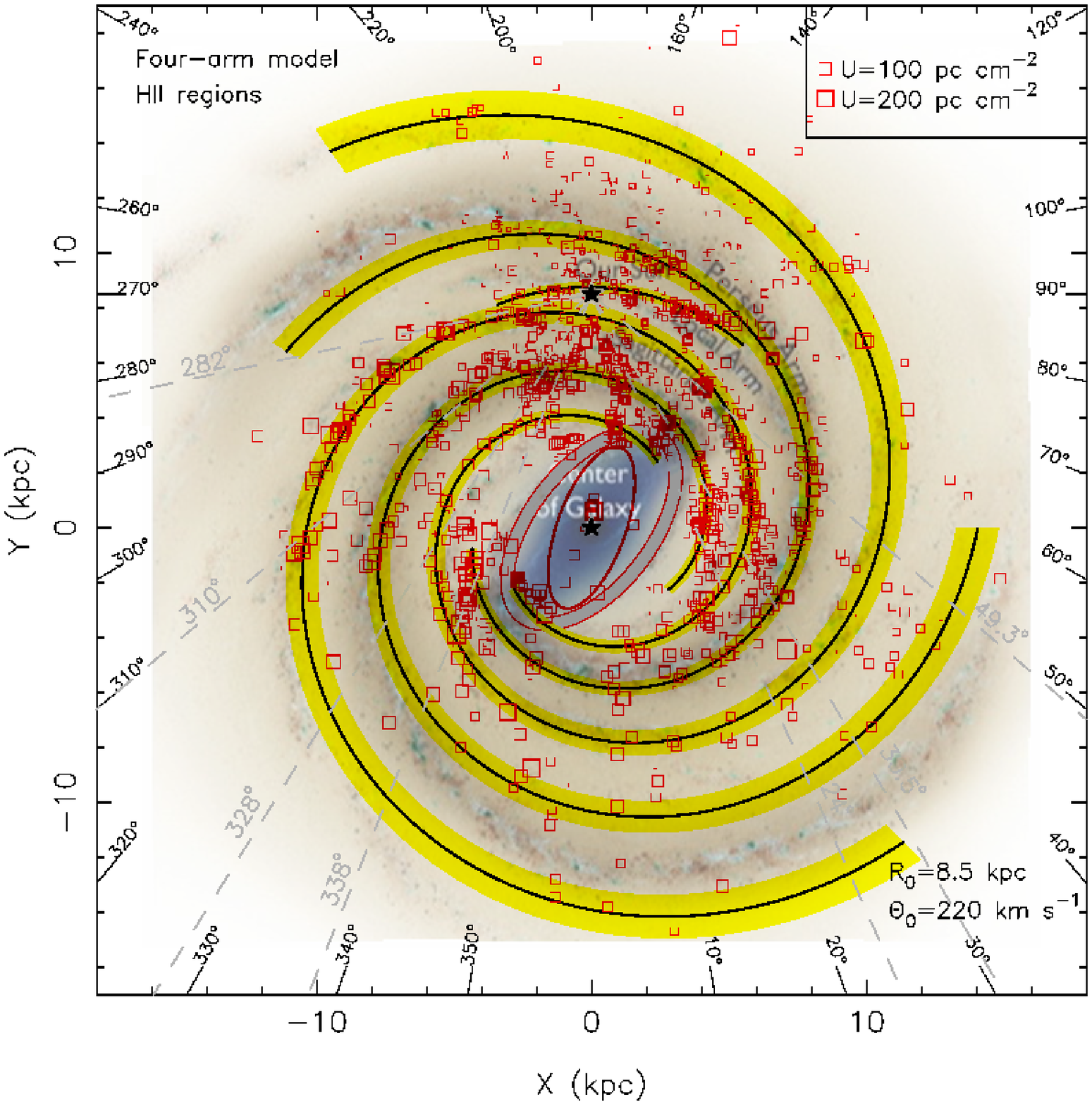}
\centering\includegraphics[width=0.47\textwidth]{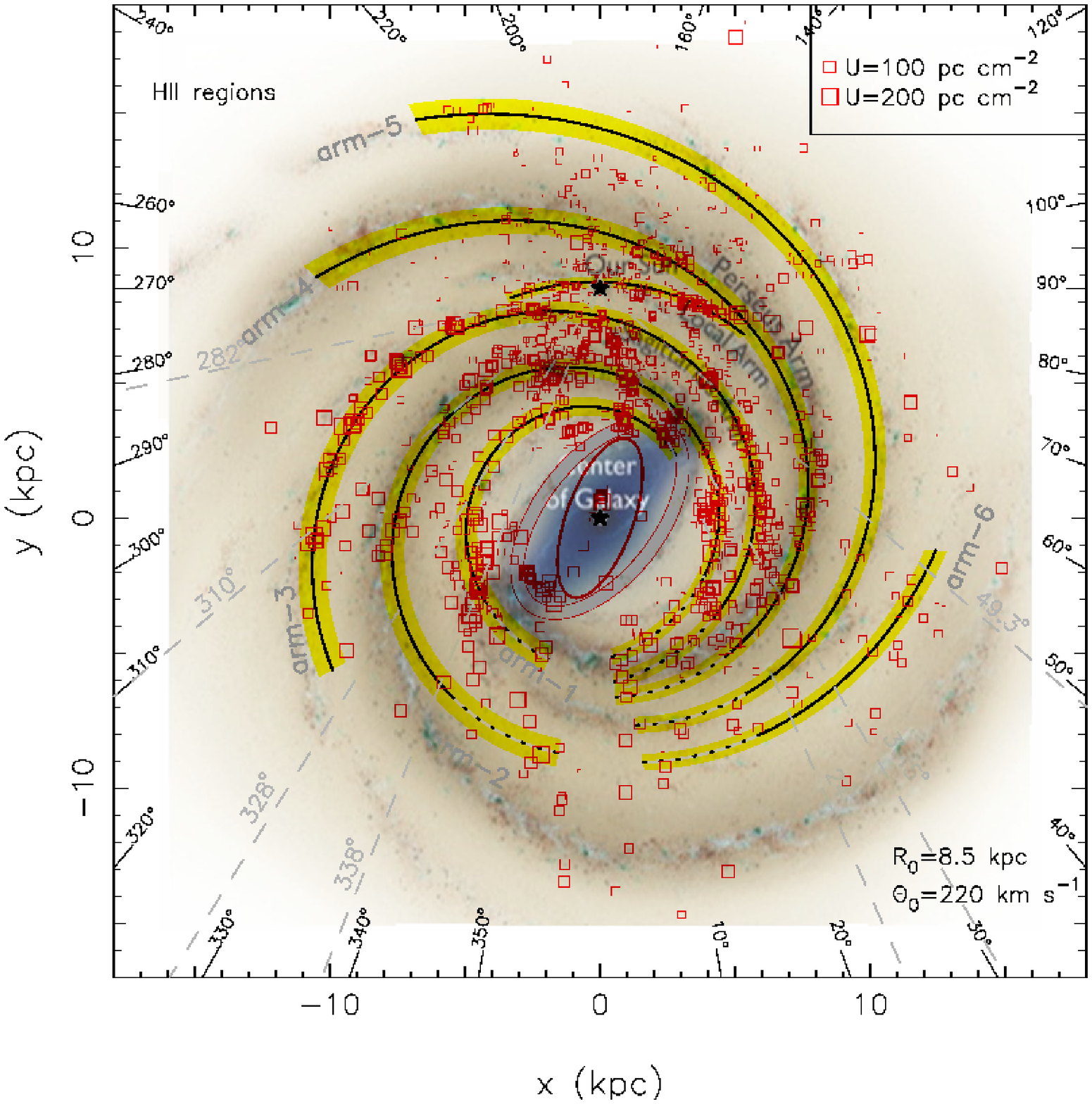}
\caption{Distribution of HII regions was overlaid on the widely used
  image of the Milky Way spiral structure
  (NASA/JPL-Caltech/R.Hurt). The symbols are the same as those in
  Fig.~\ref{lvall}. The outlines are the best-fitted four-arm model
  ({\it left}, also see the {\it upper right panel} of
  Fig.~\ref{model_log}) and the best-fitted PL spiral arm model ({\it
    right}, also see the {\it left panel} of Fig.~\ref{model_pol}).}
\label{overlay}
\end{figure*}

The best-fitted model to all three kinds of tracers is shown in the
right panel of Fig.~\ref{model_pol}. The best-fitted arm-1 to arm-4,
and also the Local Arm are consistent in the models. Large
discrepancies in positions are found for the best-fitted Outer Arm and
the Outer+1 Arm in this model compared with the model of HII regions
partially because of the lack of HII regions in the outer Galaxy and
the unreliability of the kinematic distances for many GMCs in the
second and third quadrants (Sect.~\ref{log_mod}). In the distant
regions of the first quadrant, the CO emission features revealed by
\citet{dt11} may trace a more distant arm segment even beyond the
Outer+1 Arm (also see Fig.~\ref{model_log}). These CO emission
features may be related with the extension of the Sagittarius-Carina
Arm (arm-3), but cannot be reasonably fitted by a single spiral
arm. More measurements of distant spiral tracers are necessary.

The pitch angle for a PL arm segment is expressed in the form of $
tan(\psiup_{i})= b_i + 2c_i\theta + 3d_i\theta^2$, which varies
significantly with the azimuthal angle as shown in
Fig.~\ref{pitch_a}. In some galaxies, e.g., NGC 628 and NGC 6946,
the variations of pitch angles are also obvious
\citep[][]{ccb+12,sr13}. The relative positions of spiral tracers in
our Galaxy have larger uncertainties than those in face-on
galaxies. Therefore, high quality data (e.g., the trigonometric
parallax) is necessary to uncover the properties of spiral arms in our
Galaxy.

\begin{figure*}
  \centering\includegraphics[width=0.47\textwidth]{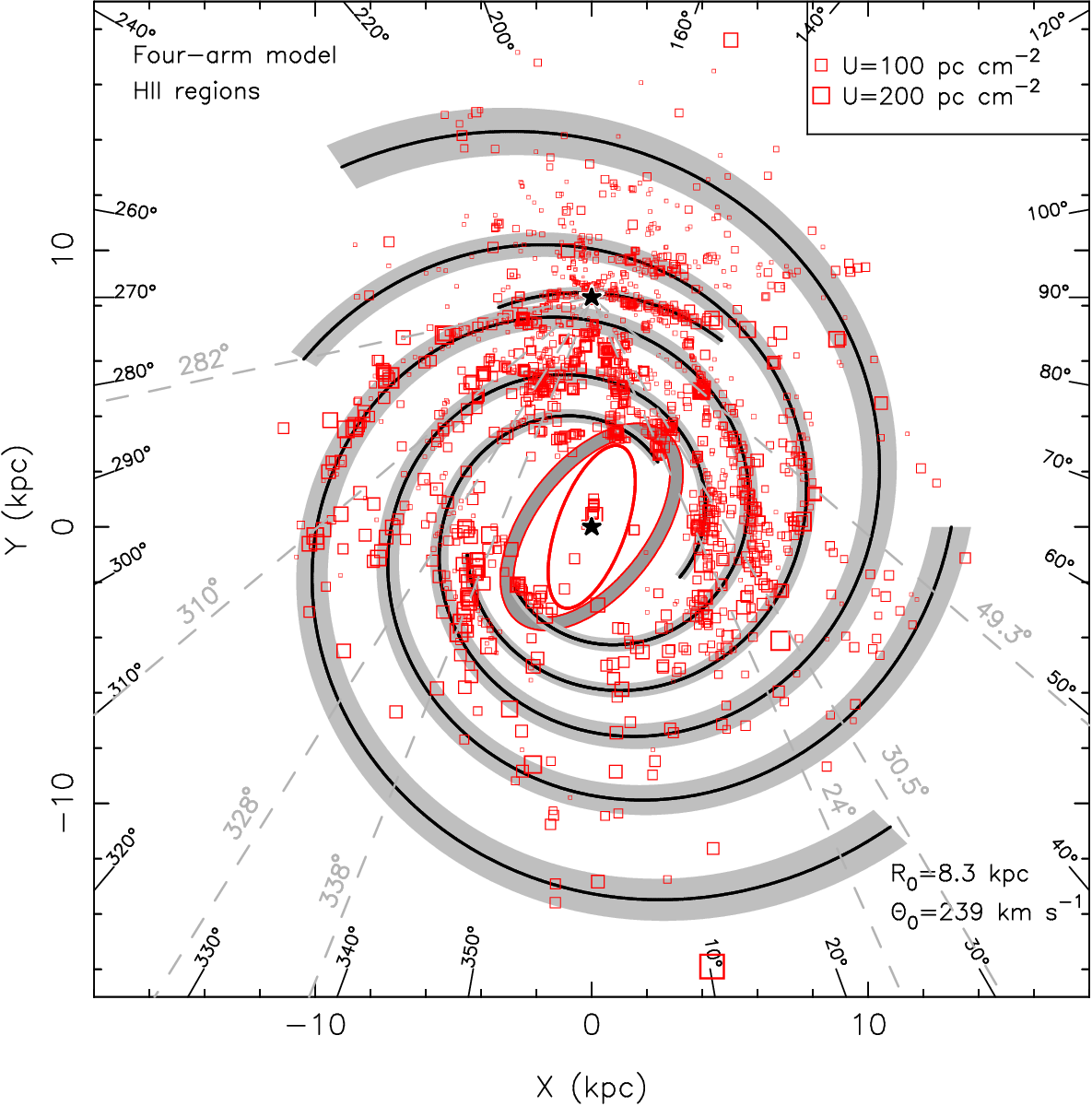}
  \centering\includegraphics[width=0.47\textwidth]{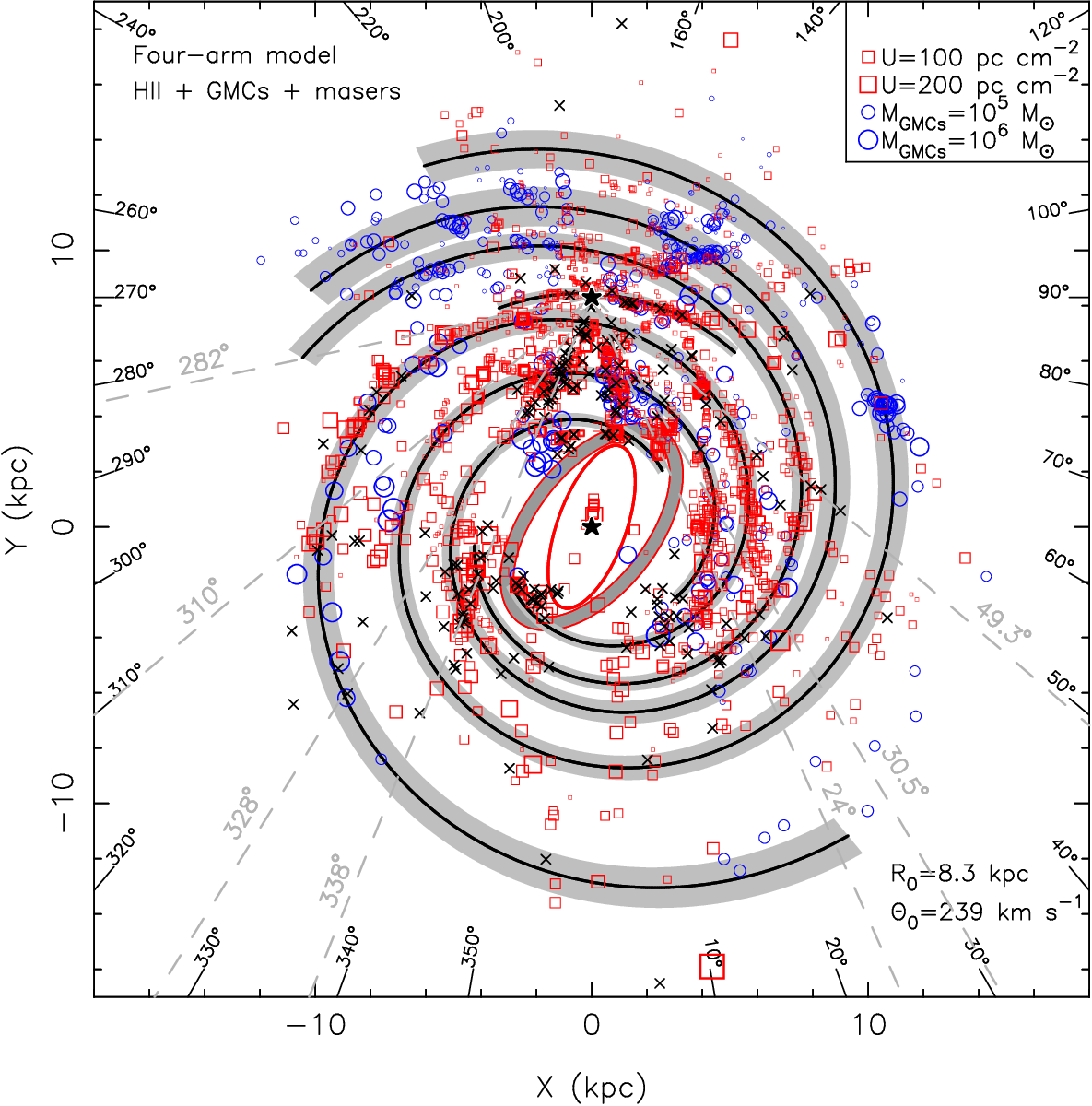}\\
  \centering\includegraphics[width=0.47\textwidth]{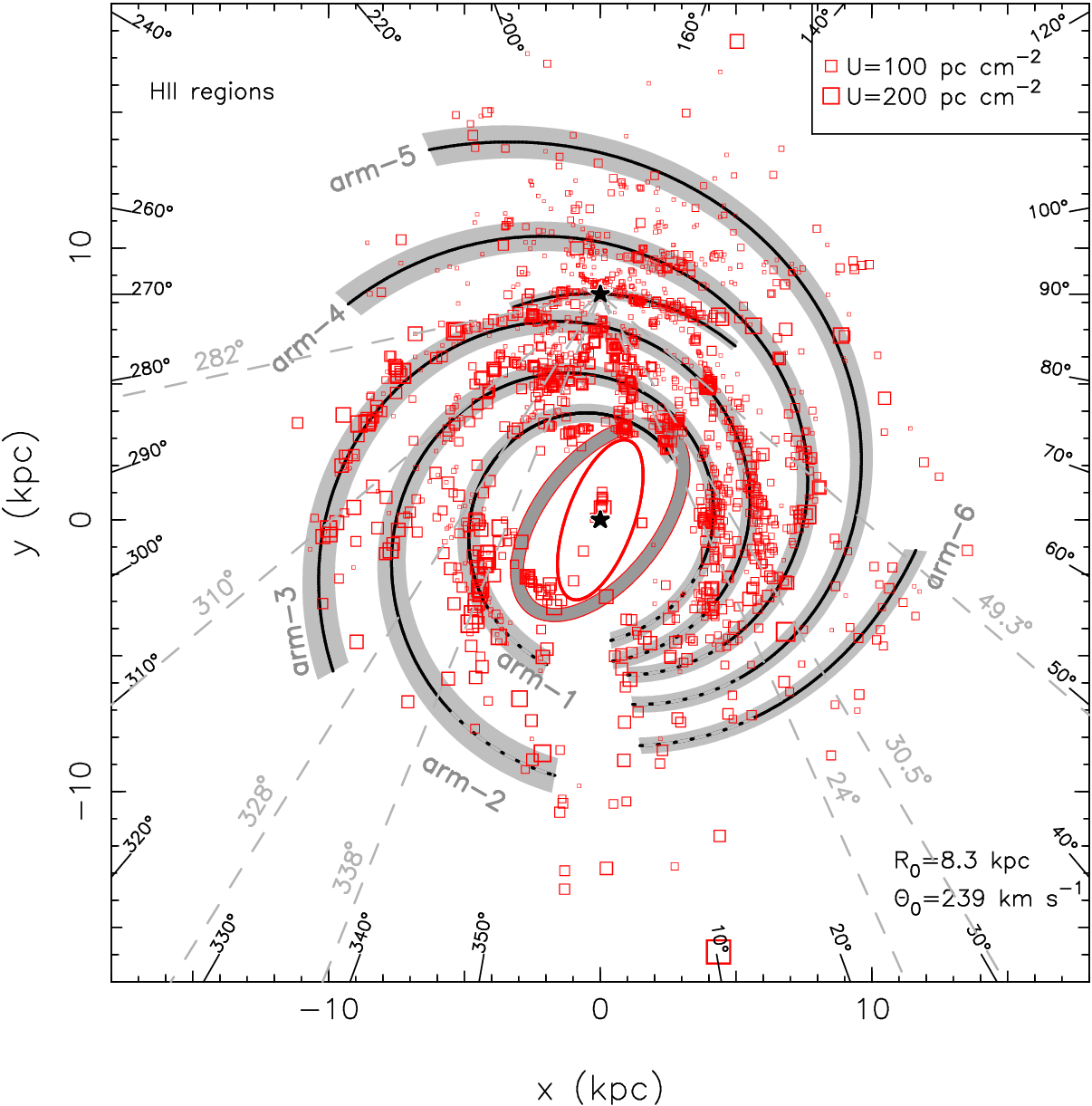}
  \centering\includegraphics[width=0.47\textwidth]{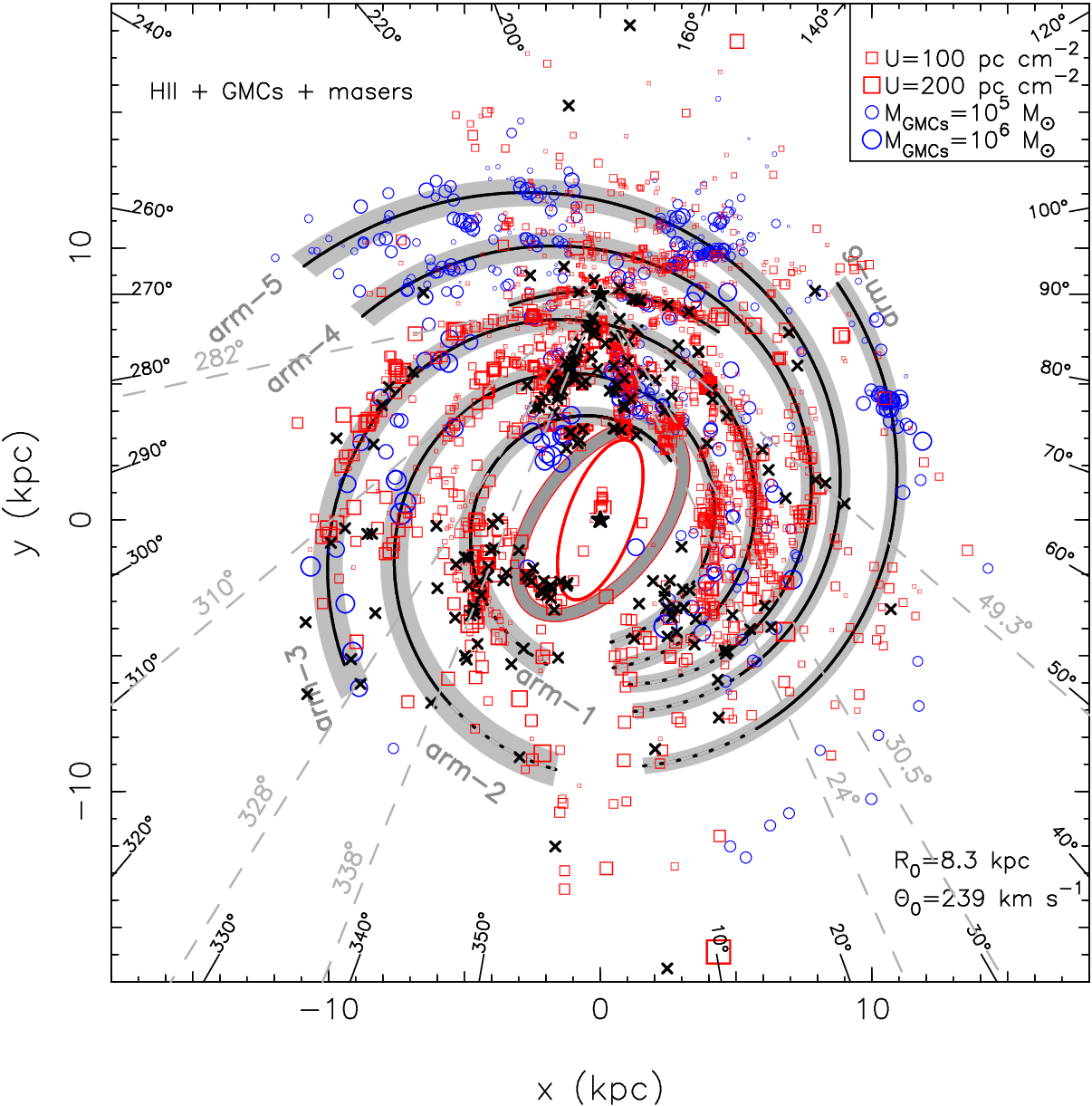}\\
  \caption{{\it Upper panels}: distributions of HII regions ({\it
      left}) and all three kinds of spiral tracers ({\it right}) with
    the flat rotation curve with newly observed $R_{0}$ = 8.3 kpc and
    $\Theta_0$ = 239 km~s$^{-1}$ \citep{brm+11}. The symbols are the
    same as those in Fig.~\ref{lvall}. The best-fitted four-arm
    models are also shown. {\it Lower panels}: the best-fitted PL
    models for these two distributions are presented.}
\label{Ro}
\end{figure*}

\subsection{Comparison with the widely used image of the Galaxy spiral
  structure}

With the updated catalogs of Galactic HII regions, GMCs, and 6.7 GHz
methanol masers, we explore the spiral pattern of the Galaxy. Some
revealed arm features are not consistent with the well-known picture
of the Galaxy spiral structure (Fig.~\ref{overlay}):

(1) The extension of the Local Arm. In the concept map of the Galaxy
structure (the background of Fig.~\ref{overlay}), the Local Arm
starts near the Perseus Arm, then extends as an independent arm
segment between the Perseus Arm and the Sagittarius-Carina Arm ($x
\sim$ $-$3 kpc, $y \sim$9 kpc). This feature is not confirmed by the
distribution of HII regions. As shown in Fig.~\ref{overlay}, the
Local Arm does not extend far from the Sun. It starts near the Perseus
Arm, then extends to the fourth quadrant, and closely approaches to
the Carina Arm at $x \sim$ $-$1.5 kpc, $y \sim$8.3 kpc, also see
Fig.~\ref{Ro}.

(2) The start point of the Scutum Arm. The Scutum Arm is commonly
believed to originate from the near end of the Galactic bar. This
contradicts the distribution of the cataloged spiral tracers. In the
best-fitted models, the Scutum Arm is related to many bright HII
regions in the distant parts of the first Galaxy quadrant around
$x\sim$4 kpc, $y\sim$0 kpc, which is consistent with the spiral arm
models of GG76 and \citet{ne2001}.

(3) The demarcation of the Sagittarius-Carina Arm and the
Scutum-Centaurus Arm. Around the direction of the GC ($|x|\lesssim$ 2
kpc, 4 kpc $\lesssim y \lesssim$ 7 kpc), no clear demarcation is found
in the data distribution between the best-fitted Sagittarius-Carina
Arm and the Scutum-Centaurus Arm, which is also shown in the
distribution of HII regions with photometric or trigonometric
distances (Fig.~\ref{dis_op_all}).

(4) The extensions of the Sagittarius-Carina Arm and the
Scutum-Centaurus Arm. As seen in Fig.~\ref{overlay}, large
discrepancies are found for the extension of the Sagittarius-Carina
Arm or the Scutum-Centaurus Arm (in the regions with $y <$ $-$6 kpc)
between our best-fitted models and that of the concept map.

\begin{figure*}
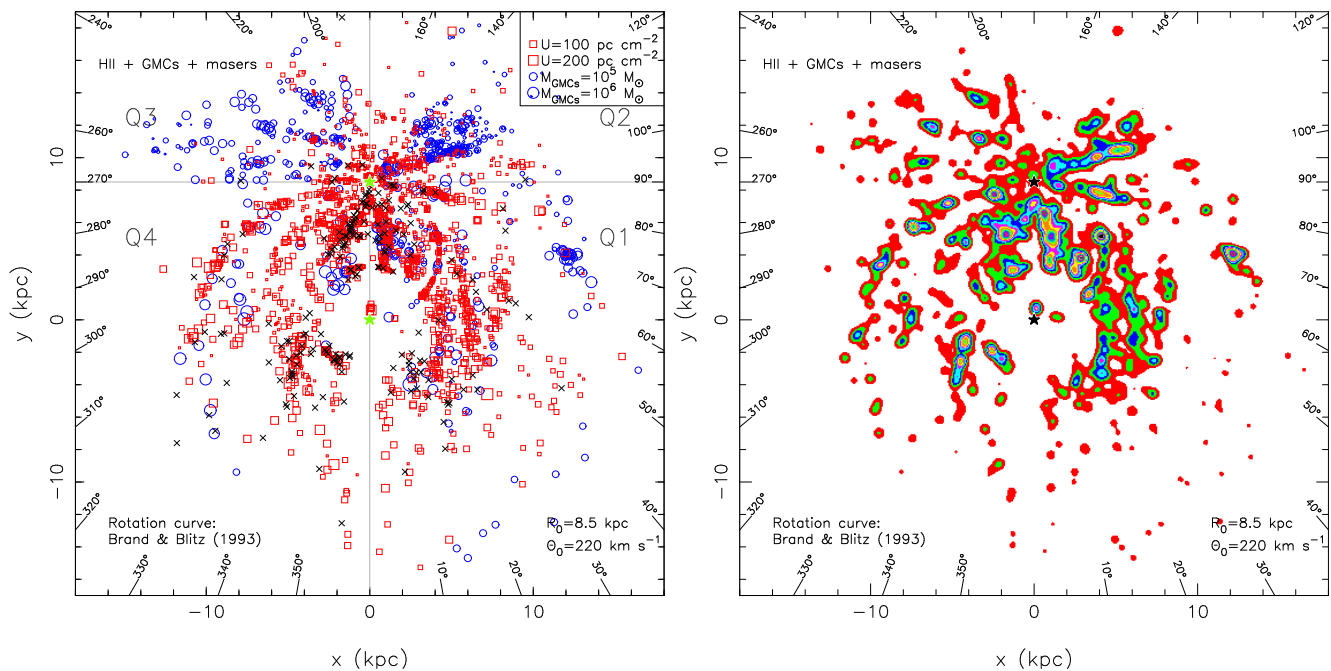

  \centering\includegraphics[width=0.47\textwidth]{dis_bb93_8.5.ps}
  \centering\includegraphics[width=0.47\textwidth]{bb93_contour_8.5.ps}\\
  \caption{{\it Left}: distributions of HII regions, GMCs, and 6.7 GHz
    methanol masers projected into the Galactic plane. The symbols are
    the same as those in Fig.~\ref{lvall}. The kinematic distances
    are estimated using the rotation curve of BB93. {\it Right}: color
    intensity map of spiral tracers. The IAU standard $R_{0}$ = 8.5
    kpc and $\Theta_0$ = 220 km~s$^{-1}$ and standard solar motions
    are adopted in deriving the kinematic distances if no photometric
    or trigonometric distance is available.}
\label{bb93}
\end{figure*}

\subsection{Influence of the fundamental parameters to the derived spiral
  structure}
\label{sec_r0}

As discussed in Sect. 2.2, recent observations suggest that $R_{0}$,
$\Theta_0$, and solar motions are different from the IAU standards,
which should influence the derived spiral pattern of our Galaxy from
the cataloged tracers. To show the possible influence, we
re-calculated the distances of spiral tracers with a flat rotation
curve for $R_{0}$ = 8.3 kpc and $\Theta_0$ = 239 km~s$^{-1}$
\citep{brm+11} together with the new solar motions of \citet{sbd10}.

The results are given in Fig.~\ref{Ro}. The arm-like features are
very similar to that with the IAU standard, but slightly shrinked,
especially for the outer Galaxy regions. We also fit the data
distribution with the four-arm logarithmic spiral arm models and the
PL arm models. The best fits are shown in Fig.~\ref{Ro}, the
corresponding arm parameters are listed in Table~\ref{34para} and
Table~\ref{5arm}. The modeled tangential directions are compared with
observations in Table~\ref{tan_mod}.

\subsection{Influence of the rotation curve on the derived spiral
  structure}
\label{sec_bb93}

Besides the flat rotation curve, the rotation curve of BB93 has also
been widely used in previous studies. As shown by \citet{hhs09}, the
BB93 rotation curve is reasonable for the whole Galaxy in comparison
with that of \citet{clem85} and \citet{fbs89}. We re-calculated the
kinematic distances of the collected spiral tracers with the BB93
rotation curve and the results are shown in Fig.~\ref{bb93}. The
arm-locations are very similar to those calculated with the flat
rotation curve. This is reasonable because the BB93 rotation curve is
almost flat in the galactocentric radii between $\sim$5 kpc and about
15 kpc.

\subsection{The Galactic warp}

The Galactic warp was first discovered in early surveys of HI gas
\citep{kerr57,burt88}, which is vertically distorted even to distances
larger than 3 kpc from the Galactic plane~\citep{ds91}. Evidence for
Galactic warp has recently been observed again from HI absorption
\citep{dsg+09}, red clump giants \citep{lch+10,boby10}, extended low
density warm ionized medium \citep{cmf+09}, 2MASS infrared stars
\citep{rmr+09}, young open clusters \citep{vmc+08}, molecular gas
\citep{nomf05}, Galactic HII regions \citep{pdd04}, and also star
forming complexes \citep{rus03}.

\begin{figure}[!b]
  \centering\includegraphics[width=0.47\textwidth]{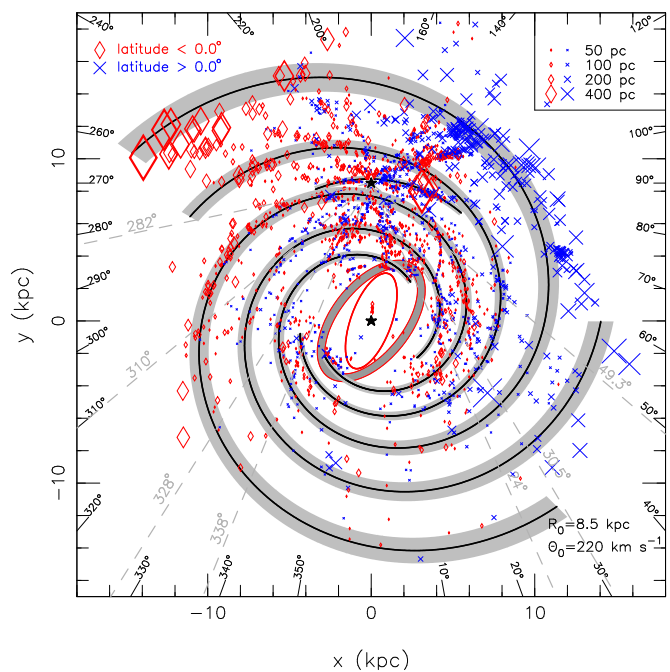}
  \caption{Evidence of Galactic warp as shown by the distributions of
    HII regions, GMCs, and 6.7 GHz methanol masers. Note that the
    diamonds here indicate the tracers of $b < 0.0^\circ$, and the
    blue crosses indicate the tracers of $b > 0.0^\circ$. The symbol
    size is proportional to the offset from the Galactic plane. The
    outlines are the best-fitted four-arm model (see the {\it upper
      right panel} of Fig.~\ref{model_log}). }
\label{warp}
\end{figure}

With the catalogs of Galactic HII regions, GMCs, and 6.7 GHz methanol
masers, we get significant evidence of warp in the outer Galaxy as
shown in Fig.~\ref{warp}, where the best-fitted four-arm model is
overlaid. In the first and second quadrants, the Outer Arm and the
Outer+1 Arm are above the Galactic plane (positive). In the third and
fourth quadrants, the Outer Arm, the Perseus Arm, and the Carina Arm
are below the Galactic plane (negative).

\section{Discussions and conclusions}

We update the catalogs of Galactic HII regions, GMCs, and 6.7 GHz
methanol masers to study the large-scale spiral arm structure.
Compared with our previous results in \citet{hhs09}, the samples of
HII regions and GMCs are much enlarged. We include the maser data
though they are marginally helpful. If the photometric or
trigonometric distances of these tracers are not available in the
literature, kinematic distances are estimated using a Galaxy rotation
curve with two sets of $R_0$, $\Theta_0$, and solar motions, where one
set is the IAU standard and the other is from the new observational
results. We calculate the excitation parameters of HII regions and the
masses of GMCs and scale them as the weight factors, then obtain the
weighted distribution of these spiral tracers and also their
combinations.

We have shown that the distribution of HII regions delineates a clear
spiral structure. Obvious arm segments can be identified. From the
inside out, they are the Norma Arm, the Scutum-Centaurus Arm, the
Sagittarius-Carina Arm, the Perseus Arm, the Outer Arm, and probably
the Outer+1 Arm. The Local Arm is identified as a short arm
segment. The identified arm segments traced by massive star forming
regions and GMCs can match the HI arms in the outer Galaxy, suggesting
that the Milky Way is a grand-design spiral galaxy.

We fit the models of three-arm and four-arm logarithmic spirals to
only the HII regions and to all three kinds of spiral tracers,
respectively. Both models can connect most spiral tracers, but the
connections of known arm segments differ in the two models. A PL
spiral arm model is able to not only connect most spiral tracers, but
also match the observed tangential directions.

\begin{acknowledgements}
  We thank the referee for helpful comments, and Dr. L{\'e}pine and
  Dr. Bronfman for kindly providing us their data. The authors are
  supported by the Strategic Priority Research Program ``The Emergence
  of Cosmological Structures" of the Chinese Academy of Sciences,
  Grant No. XDB09010200, and the National Natural Science Foundation (NNSF) of
  China No. 11473034. L.G.H. is also supported by the Young Researcher
  Grant of National Astronomical Observatories, Chinese Academy of
  Sciences.
\end{acknowledgements}

\scriptsize
\bibliographystyle{aa} 
\bibliography{Galac_structure}


\normalsize
\begin{appendix}
\section{Tracer data for spiral structure}
The related parameters of HII regions, GMCs, and 6.7 GHz methanol
masers have been collected and calculated based on the data in
literature, as listed in Table~\ref{tab_a1}$-$\ref{tab_a3},
respectively.

\newcommand{\wuhao}{\fontsize{5.5pt}{5.5pt}\selectfont}

\wuhao
\setlength{\tabcolsep}{0.13mm}
\begin{longtable}{llrrrrrclllllrrrllccccccccc}
  \caption{\label{tab_a1} HII regions as tracers of spiral arms. \\
    {\bf Notes.} A
    portion of this table is listed here. The entire version will be
    available as a machine-readable form at the CDS and at the
    authors' webpage. Columns 1$-$3 list the Number, the Galactic
    longitude, and the Galactic latitude for HII regions, taken from
    the reference listed in Column 4; Column 5 and 6 list the velocity
    and its uncertainty if available. The $V_{\rm LSR}$ are directly
    taken from references. The standard solar motion $[$20.0 km/s,
      toward R.A.(1900) = 18h and Decl. (1900) = 30 deg$]$ was
    generally adopted to get $V_{\rm LSR}$ ; Column 7 is the measured
    electron temperature; Column 8 and 9 list the used telescope and
    the observed line(s) derived from the reference in Column 10. The
    abbreviations for telescopes are: GBT--the Green Bank 100~m
    telescope; Are--the Arecibo 305~m telescope; B\&M--the 7~m
    telescope of the Bell Telescope Laboratories and the 5~m telescope
    at the Millimeter Wave Observatory; SET--the Swedish-ESO
    Submillimeter telescope; OSO--the Onsala Space Observatory 20~m
    telescope; PKS--the Parkes 64~m telescope; GBO--the 42.6~m
    telescope of the NRAO at Green Bank; BRAL--the 25.8~m telescope of
    the Berkeley Radio Astronomy Laboratory; Eff--the Effelsberg 100~m
    telescope; XAO--the 25~m telescope of Xinjiang Observatory;
    LO--the 0.76~m telescope at the Leuschner Observatory; DRAO--The
    25.6~m telescope at the Dominion Radio Astrophysical Observatory;
    ARO--the 46~m telescope of the Algonquin Radio Observatory. Column
    11$-$13 are the radio continuum flux of the HII region, the
    observation frequency and the reference if available. Column
    14$-$16 list the photometric or trigonometric distance when
    available, as well as the uncertainty and the reference; Column 17
    is a note for the kinematic distance ambiguity and Column 18 is
    the reference(s) and note: knear--the nearer kinematic distance is
    adopted; kfar--the farther kinematic distance is adopted;
    ktan--the tracer is located at the tangential point and the
    distance to the tangent is adopted; kout--the tracer is in the
    outer Galaxy; n3kpc--the tracer is located in the Near 3 kpc Arm;
    f3kpc--the tracer is in the Far 3 kpc Arm; Columns 19 and 20 list
    the distance to the GC and the distance to the Sun, the kinematic
    distances when adopted are calculated with a flat rotation curve
    with the IAU standard $R_{0}$ = 8.5 kpc and $\Theta_0$ = 220
    km~s$^{-1}$ and the standard solar motions; Column 21 is the
    excitation parameter re-scaled by the adopted distance; Columns
    22$-$24 are the same as Columns 19$-$21, except with fundamental
    parameters $R_{0}$ = 8.3 kpc and $\Theta_0$ = 239 km~s$^{-1}$
    \citep{brm+11} and the solar motions of \citet{sbd10}; Columns
    25$-$27 are the same as Columns 19$-$21, except with the rotation
    curve of \citet{bb93}.\\
    {\bf References.} ab09: \citet{ab09}; abbr11: \citet{abbr11}; abbr12:
    \citet{abbr12}; adps79: \citet{adps79}; adg+70: \citet{adg+70};
    ahck02: \citet{ahck02}; alb80: \citet{alb80}; bab12:
    \citet{bab12}; bfs82: \citet{bfs82}; bnm96: \citet{bnm96}; brba11:
    \citet{brba11}; brm+09: \citet{brm+09}; bwl09: \citet{bwl09};
    casw72: \citet{casw72}; cf08: \citet{cf08}; ch87: \citet{ch87};
    cmh74: \citet{cmh74}; cor+07: \citet{cor+07}; cpt06:
    \citet{cpt06}; diet67: \citet{diet67}; dlr+11: \citet{dlr+11};
    dm72: \citet{dm72}; dsa+10: \citet{dsa+10}; dwbw80:
    \citet{dwbw80}; dze+11: \citet{dze+11}; fb84: \citet{fb84}; fb14:
    \citet{fb14}; fc72: \citet{fc72}; frwc03: \citet{frwc03}; fdt90:
    \citet{fdt90}; F3R: \citet{frrr90}; GB6: \citet{gsdc96}; gg76:
    \citet{gg76}; gm11: \citet{gm11}; gmb+07: \citet{gmb+07}; hab+08:
    \citet{hab+08}; hbc+07: \citet{hbc+07}; hbc+08: \citet{hbc+08};
    hbc+07b: \citet{hbc+07b}; hbm+09: \citet{hbm+09}; hls+00:
    \citet{hls+00}; hze+11: \citet{hze+11}; hhi+11: \citet{hhi+11};
    irm+13: \citet{irm+13}; jd12: \citet{jd12}; kb94: \citet{kb94};
    kc97: \citet{kc97}; kcw94: \citet{kcw94}; kjb+03: \citet{kjb+03};
    kns+11: \citet{kns+11}; kwc77: \citet{kwc77}; lock79:
    \citet{lock79}; lock89: \citet{lock89}; lph96: \citet{lph96};
    lra11: \citet{lra+11}; lt08: \citet{lt08}; lbbs14: \citet{lbbs14};
    mcr+11: \citet{mcr+11}; mdf+11: \citet{mdf+11}; mg68:
    \citet{mg68}; mgps2: \citet{mmg+07}; mh67: \citet{mh67}; mitg:
    \citet{blb+86}; mn81: \citet{mn81}; mnb75: \citet{mnb75}; mrm+09:
    \citet{mrm+09}; mcg09: \citet{mcg09}; msh+11: \citet{msh+11};
    nnh+11: \citet{nnh+11}; okh+10: \citet{okh+10}; pbd+03:
    \citet{pbd+03}; pbt77: \citet{pbt77}; pco10: \citet{pco10}; pdd04:
    \citet{pdd04}; pedl80: \citet{pedl80}; PMN:
    \citet{wgbe94,wgh+96,gwbe94,gwbe95}; qrbb06: \citet{qrbb06};
    rag07: \citet{rag07}; rbr+10: \citet{rbr+10}; rbs+12:
    \citet{rbs+12}; rmb+09: \citet{rmb+09}; rmhg80: \citet{rmhg80};
    rmz+09: \citet{rmz+09}; ross78: \citet{ross78}; rus03:
    \citet{rus03}; rwb+70: \citet{rwb+70}; shd+11: \citet{shd+11};
    shr+10: \citet{shr+10}; srbm10: \citet{srbm10}; srd+12:
    \citet{srd+12}; srm+09: \citet{srm+09}; srm+14: \citet{srm+14};
    st79: \citet{st79}; swa+04: \citet{swa+04}; uhl12: \citet{uhl12};
    vch76: \citet{vch76}; vgd+10: \citet{vgd+10}; wa72: \citet{wa72};
    wam82: \citet{wam82}; was+03: \citet{was+03}; wend70:
    \citet{wend70}; wmgm70: \citet{wmgm70}; wsr+14: \citet{wsr+14};
    wwb83: \citet{wwb83}; wxm+12: \citet{wxm+12}; xlr+13:
    \citet{xlr+13}; xmr+11: \citet{xmr+11}; xrm+08: \citet{xrm+08};
    xrm+09: \citet{xrm+09}; xrzm06: \citet{xrzm06}; zms+14:
    \citet{zms+14}; zrm+13: \citet{zrm+13}; zzr+09: \citet{zzr+09};
    87GB: \citet{gc91}.  }  \\ \hline \hline No. & Glon & Glat & Ref. &
  $V_{\rm LSR}$ & $\sigma$$_{\rm VLSR}$ & T$_e$ & Tel. & lines &
  Ref. & S & Freq. & Ref. & D & $\sigma_{\rm D}$ & Ref. & Mark &
  Ref.\& Note & R$_{8.5}$ & D$_{8.5}$ & U$_{8.5}$ & R$_{8.3}$ &
  D$_{8.3}$ & U$_{8.3}$ & R$_{\rm BB93}$ & D$_{\rm BB93}$ & U$_{\rm
    BB93}$ \\ \\ &$(^\circ)$& $(^\circ)$& & km~s$^{-1}$ & km~s$^{-1}$
  & K & & & & Jy & GHz & & kpc & kpc & & & & kpc & kpc & pc~cm$^{-2}$
  & kpc & kpc & pc~cm$^{-2}$ & kpc & kpc & pc~cm$^{-2}$ \\ (1) & (2)
  &(3) & (4) & (5) & (6) & (7) & (8) & (9) & (10)& (11) & (12) & (13)
  & (14)& (15) &(16) & (17) & (18) & (19) & (20) & (21) & (22) & (23)
  & (24) & (25) & (26) & (27) \\ \hline \endfirsthead
\caption{continued.}\\
\hline\hline
No. & Glon
  & Glat & Ref. & $V_{\rm LSR}$ & $\sigma$$_{\rm VLSR}$ & T$_e$ &
  Tel. & lines & Ref. & S & Freq. & Ref. & D & $\sigma_{\rm D}$ &
  Ref. & Mark & Ref.\& Note & R$_{8.5}$ & D$_{8.5}$ & U$_{8.5}$ &
  R$_{8.3}$ & D$_{8.3}$ & U$_{8.3}$ & R$_{\rm BB93}$ & D$_{\rm BB93}$
  & U$_{\rm BB93}$ \\ \\ &$(^\circ)$& $(^\circ)$& & km~s$^{-1}$ & km~s$^{-1}$ & K &
  & & & Jy & GHz & & kpc & kpc & & & & kpc & kpc & pc~cm$^{-2}$ & kpc
  & kpc & pc~cm$^{-2}$ & kpc & kpc & pc~cm$^{-2}$ \\ (1) & (2) &(3) &
  (4) & (5) & (6) & (7) & (8) & (9) & (10)& (11) & (12) & (13) & (14)&
  (15) &(16) & (17) & (18) & (19) & (20) & (21) & (22) & (23) & (24) &
  (25) & (26) & (27)
\\
\hline
\endhead
\hline
\endfoot
     1 &   0.02  &   0.13   &  bfs82&   $-$5.5  &  1.0 &         &Leu &  H$\alpha$    &fdt90  &         &       &       &  1.5  &  0.2   & rus03  & kfar   & rus03,S17 bfs82,G359.5$-$0.6 rus03   &   7.00&    1.50&        &   6.80&    1.50&        &   7.00&    1.50&         \\
     2 &   0.09  &   0.01   &  wwb83&  $-$25.4  &  1.9 &  6600   &Eff &  H76$\alpha$  &wwb83  &         &       &       &       &        &        &        &      ,close to GC wwb83            &       &        &        &       &        &        &       &        &         \\
       &   0.068 &   0.014  & dwbw80&  $-$34.0  &  5.0 &  4800   &Eff &  H110$\alpha$ &dwbw80 &  53.3   & 4.8   &dwbw80 &       &        &        &        &                                    &       &        &        &       &        &        &       &        &         \\
     3 &   0.094 &  $-$0.154  &  bnm96&   16.0  &      &         &SET &  CS(2$-$1)    &bnm96  &         &       &       &       &        &        &        &      ,17432$-$2855 bnm96             &       &        &        &       &        &        &       &        &         \\
     4 &   0.11  &  $-$0.56   &  bfs82&   18.1  &  0.9 &         &B\&M &  CO(1$-$0)    &bfs82  &         &       &       &  1.5  &  0.2   & rus03  & kfar   & rus03,S19 bfs82,G359.5$-$0.6 rus03   &   7.00&    1.50&        &   6.80&    1.50&        &   7.00&    1.50&         \\
       &         &          &       &   11.6  &  2.4 &         &Leu &  H$\alpha$    &fdt90  &         &       &       &       &        &        &        &      ,S19 fdt90                    &       &        &        &       &        &        &       &        &         \\
     5 &   0.18  &  $-$0.05   &  wwb83&   24.5  &  0.2 &  8800   &Eff &  H76$\alpha$  &wwb83  &         &       &       &       &        &        &        &      ,close to GC wwb83            &       &        &        &       &        &        &       &        &         \\
       &   0.178 &  $-$0.050  & dwbw80&   32.0  &  5.0 & 15900   &Eff &  H110$\alpha$ &dwbw80 & 196.9   & 4.8   &dwbw80 &       &        &        &        &                                    &       &        &        &       &        &        &       &        &         \\
     6 &   0.2   &  $-$0.1    &   mh67&   59.4  &  3.0 &         &GBO &  H109$\alpha$ &mh67   &         &       &       &       &        &        &        &      ,SgrA mh67                    &       &        &        &       &        &        &       &        &         \\
     7 &   0.2   &   0.0    & wmgm70&  $-$12.7  &  3.5 &  7600   &PKS &  H109$\alpha$ &wmgm70 & 180.0   & 5.0   &wmgm70 &       &        &        &        &      ,near GC wmgm70               &       &        &        &       &        &        &       &        &         \\
     8 &   0.284 &  $-$0.478  & dwbw80&   20.0  &  5.0 & 13000   &Eff &  H110$\alpha$ &dwbw80 &   2.2   & 4.8   &dwbw80 &       &        &        & knear  & rus03                              &   0.46&    8.05&   75.65&   0.45&    7.85&   74.44&   0.41&    8.09&   75.93 \\
       &   0.279 &  $-$0.484  &  bnm96&   18.5  &      &         &SET &  CS(2$-$1)    &bnm96  &         &       &       &       &        &        &        &      ,17449$-$2855 bnm96             &       &        &        &       &        &        &       &        &         \\
     9 &   0.33  &  $-$0.19   &  bfs82&   19.2  &  0.4 &         &B\&M &  CO(1$-$0)    &bfs82  &   0.150 & 0.843 & mgps2 &  1.5  &  0.2   & rus03  & kfar   & rus03,S20 bfs82,G359.5$-$0.6 rus03   &   7.00&    1.50&    8.78&   6.80&    1.50&    8.78&   7.00&    1.50&    8.78 \\
       &         &          &       &    9.6  &  0.5 &         &Leu &  H$\alpha$    &fdt90  &         &       &       &       &        &        &        &      ,S20 fdt90                    &       &        &        &       &        &        &       &        &         \\
       &   0.314 &  $-$0.199  &  bnm96&   18.7  &      &         &SET &  CS(2$-$1)    &bnm96  &         &       &       &       &        &        &        &      ,17439$-$2845 bnm96             &       &        &        &       &        &        &       &        &         \\
    10 &   0.361 &  $-$0.780  & dwbw80&   20.0  &  5.0 &  5100   &Eff &  H110$\alpha$ &dwbw80 &   3.8   & 4.8   &dwbw80 &       &        &        & knear  & rus03                              &   0.55&    7.95&   80.75&   0.54&    7.76&   79.46&   0.50&    8.00&   81.10 \\
    ... &   ... &  ...  & ... &  ...  &  ... &  ...   &... &  ... &... &   ...   & ...   &...&  ...     &  ...      & ...       & ...  & ...                              &   ...&    ...&   ...&   ...&    ...&   ...&   ...&    ...&   ... \\

\hline\hline
\end{longtable}

\wuhao
\setlength{\tabcolsep}{0.2mm}
\begin{longtable}{rrcrrrrrrrrllccccccccc}
  \caption{\label{tab_a2} GMCs as tracers of spiral arms. \\
    {\bf Notes.} A portion of
    this table is shown here. The entire version will be available as
    a machine-readable form at the CDS and at the authors'
    webpage. Columns 1 and 2 are the Galactic longitude and the
    Galactic latitude; Column 3 to 6 list the velocity, the luminosity
    of CO emission line, the distance and mass of the molecular cloud,
    which are obtained from the reference listed in Column 13; Column
    7 and 8 list the size of the cloud in the directions of Galactic
    longitude and latitude; Column 9 to 11 list the photometric or
    trigonometric distance, its uncertainty and the reference when
    available; Column 12 is the note for the kinematic distance
    ambiguity, and the reference is in Column 13; Columns 14 and 15
    give the distance to the GC and the distance to the Sun, the
    kinematic distances when adopted are calculated with a flat
    rotation curve with the IAU standard $R_{0}$ = 8.5 kpc and
    $\Theta_0$ = 220 km~s$^{-1}$ and the standard solar motions;
    Column 16 is the GMC mass re-scaled by the adopted distance;
    Columns 17$-$19 are the same as Columns 14$-$16, except with
    fundamental parameters $R_{0}$ = 8.3 kpc and $\Theta_0$ = 239
    km~s$^{-1}$ \citep{brm+11} and the solar motions of \citet{sbd10};
    Columns 20$-$22 are the same as Columns 14$-$16, except with the
    rotation curve of \citet{bb93}. \\
{\bf References.} bw94: \citet{bw94}; bnt89: \citet{bnt89}; css90: \citet{css90}; cgm+85: \citet{cgm+85}; dt11: \citet{dt11}; dkm+08: \citet{dkm+08}; dgt94: \citet{dgt94}; dtb90: \citet{dtb90}; fb14: \citet{fb14}; gbnd14: \citet{gbnd14}; gcbt88: \citet{gcbt88}; hcs01: \citet{hcs01}; hbc+07b: \citet{hbc+07b}; mab97: \citet{mab97}; mk88: \citet{mk88}; nomf05: \citet{nomf05}; ntbc87: \citet{ntbc87}; nbt89: \citet{nbt89}; rmb+09: \citet{rmb+09}; rjh+09: \citet{rjh+09}; sod91: \citet{sod91}; srby87: \citet{srby87}; sh10: \citet{sh10}.
} \\
\hline \hline
Glon & Glat & $V_{\rm LSR}$& $L_{\rm CO}$ &D & $M_{\rm GMCs}$ &dlon & dlat  & D & $\sigma_{\rm D}$ & Ref. & Mark & Ref. & R$_{8.5}$ & D$_{8.5}$  & M$_{8.5}$ & R$_{8.3}$  & D$_{8.3}$  &M$_{8.3}$ & R$_{\rm BB93}$ & D$_{\rm BB93}$ & M$_{\rm BB93}$\\
\\
$(^\circ)$& $(^\circ)$& km s$^{-1}$& $10^3$K~km~s$^{-1}$~pc$^2$ & kpc& $10^5M_{\odot}$& $(^\circ)$  & $(^\circ)$ & kpc & kpc & & & & kpc & kpc & $10^5M_{\odot}$ & kpc & kpc & $10^5M_{\odot}$ & kpc & kpc & $10^5M_{\odot}$ \\
  (1) & (2) &(3) & (4) & (5)  & (6)  & (7) & (8) & (9) & (10) & (11) & (12) & (13)& (14) & (15) & (16) & (17) & (18) & (19) & (20) & (21) & (22)\\
\hline
\endfirsthead
\caption{continued.}\\
\hline\hline
Glon & Glat & $V_{\rm LSR}$& $L_{\rm CO}$ &D & $M_{\rm GMCs}$ &dlon & dlat  & D & $\sigma_{\rm D}$ & Ref. & Mark & Ref. & R$_{8.5}$ & D$_{8.5}$  & M$_{8.5}$ & R$_{8.3}$  & D$_{8.3}$  &M$_{8.3}$ & R$_{\rm BB93}$ & D$_{\rm BB93}$ & M$_{\rm BB93}$\\
\\
$(^\circ)$& $(^\circ)$& km s$^{-1}$& $10^3$K~km~s$^{-1}$~pc$^2$ & kpc& $10^5M_{\odot}$& $(^\circ)$  & $(^\circ)$ & kpc & kpc & & & & kpc & kpc & $10^5M_{\odot}$ & kpc & kpc & $10^5M_{\odot}$ & kpc & kpc & $10^5M_{\odot}$ \\
  (1) & (2) &(3) & (4) & (5)  & (6)  & (7) & (8) & (9) & (10) & (11) & (12) & (13)& (14) & (15) & (16) & (17) & (18) & (19) & (20) & (21) & (22)\\
\hline
\endhead
\hline
\endfoot
   8.00  &  $-$0.50  &          128.  &    72.7 &  10.1 &   4.44&  0.06 &  0.07 &         &        &         & ktan & srby87                        &   1.64&    9.55&    3.97&   1.65&    9.40&    3.84&   1.18&    8.42&    3.08 \\
   8.20  &   0.20  &           20.  &  1402.0 &  15.9 &  17.63&  0.17 &  0.21 &         &        &         & kfar & srby87                        &   5.19&   13.46&   12.64&   5.02&   13.09&   11.96&   5.17&   13.44&   12.59 \\
   8.30  &   0.00  &            3.  &   226.0 &   6.2 &   6.54&  0.40 &  0.11 &         &        &         & 3kpc & srby87                        &       &        &        &       &        &        &       &        &         \\
   8.30  &  $-$0.10  &           48.  &    50.2 &  13.2 &   1.11&  0.05 &  0.05 &         &        &         & kfar & srby87                        &   3.38&   11.57&    0.85&   3.34&   11.33&    0.82&   3.31&   11.49&    0.84 \\
   8.30  &  $-$0.30  &           16.  &    28.1 &   3.2 &   1.00&  0.16 &  0.15 &         &        &         & knear& srby87                        &   5.65&    2.89&    0.82&   5.44&    2.91&    0.83&   5.64&    2.90&    0.82 \\
   8.40  &  $-$0.30  &           37.  &   233.0 &   5.7 &   6.65&  0.32 &  0.15 &         &        &         & knear& srby87                        &   3.95&    4.66&    4.44&   3.87&    4.53&    4.20&   3.89&    4.72&    4.57 \\
   8.50  &  $-$1.00  &           16.  &   106.0 &   3.2 &   3.37&  0.25 &  0.25 &         &        &         & knear& srby87                        &   5.70&    2.85&    2.67&   5.48&    2.87&    2.70&   5.69&    2.86&    2.68 \\
   8.70  &   0.60  &           22.  &     8.6 &   4.0 &   0.42&  0.08 &  0.09 &         &        &         & knear& srby87                        &   5.12&    3.45&    0.31&   4.96&    3.41&    0.30&   5.09&    3.48&    0.32 \\
   8.90  &  $-$0.50  &           12.  &     9.4 &   6.3 &   0.84&  0.04 &  0.04 &         &        &         & 3kpc & srby87                        &       &        &        &       &        &        &       &        &         \\
   ...  &   ...  &        ...  &    ... &   ... &   ...&  ... &  ... &  ...    &  ...    &   ...    & ...& ...                        &   ...&    ...&    ...&   ...&    ...&    ...&   ...&    ...&    ... \\

\hline\hline
\end{longtable}

\wuhao
\setlength{\tabcolsep}{1.0mm}
\begin{longtable}{lrrrrrrrllllllll}
  \caption{\label{tab_a3} 6.7 GHz methanol masers as tracers of spiral
    arms. \\
    {\bf Notes.} A portion of this table is shown here. The entire version
    will be available as a machine-readable form at the CDS and at the
    authors' webpage. Columns 1 and 2 are the Galactic longitude and
    latitude; Column 3 and 4 list the median velocity and the peak
    velocity of the maser lines, the reference is listed in Column 5;
    Column 6$-$8 list the photometric or trigonometric distance, its
    uncertainty and the reference when available; Column 9 is the note
    for the kinematic distance ambiguity; the reference is shown in
    Column 10; Columns 11 and 12 list the distance to the GC and the
    distance to the Sun, the kinematic distances when adopted are
    calculated with a flat rotation curve with the IAU standard
    $R_{0}$ = 8.5 kpc and $\Theta_0$ = 220 km~s$^{-1}$ and the
    standard solar motions; Columns 13 and 14 are the same as Columns
    11 and 12 except with fundamental parameters $R_{0}$ = 8.3 kpc and
    $\Theta_0$ = 239 km~s$^{-1}$ \citep{brm+11} and the solar motions
    of \citet{sbd10}; Columns 15 and 16 are the same as Columns 11 and
    12, except with the rotation curve of \citet{bb93}.\\
    {\bf References.} ab09: \citet{ab09}; bdc01: \citet{bdc01}; bbs+08:
    \citet{bbs+08} brm+09: \citet{brm+09}; bph+06: \citet{bph+06};
    cmr+75: \citet{cmr+75}; chr+14: \citet{chr+14}; dlr+11:
    \citet{dlr+11}; dwbw80: \citet{dwbw80}; gcf+12: \citet{gcf+12};
    gm11: \citet{gm11}; hbc+07: \citet{hbc+07}; hbc+07b:
    \citet{hbc+07b}; hhk+11: \citet{hhk+11}; irm+13: \citet{irm+13};
    kns+11: \citet{kns+11}; kjb+03: \citet{kjb+03}; mrm+09:
    \citet{mrm+09}; msh+11: \citet{msh+11}; mdf+11: \citet{mdf+11};
    mcr+11: \citet{mcr+11}; mdf+11: \citet{mdf+11}; nnh+11:
    \citet{nnh+11}; okh+10: \citet{okh+10}; oah+13: \citet{oah+13};
    pmg08: \citet{pmg08}; pmg09: \citet{pmg09}; pmb05:
    \citet{pmb05}; rjh+09: \citet{rjh+09}; rbr+10: \citet{rbr+10};
    rbs+12: \citet{rbs+12}; rmb+09: \citet{rmb+09}; swa+04:
    \citet{swa+04}; syby87: \citet{srby87}; srm+09: \citet{srm+09};
    shr+10: \citet{shr+10}; sxc+14: \citet{sxc+14}; sh10:
    \citet{sh10}; uhl12: \citet{uhl12}; wam82: \citet{wam82}; wxm+12:
    \citet{wxm+12}; wsr+14: \citet{wsr+14}; xrzm06: \citet{xrzm06};
    xlh+08: \citet{xlh+08}; xrm+09: \citet{xrm+09}; xmr+11:
    \citet{xmr+11}; xlr+13: \citet{xlr+13}; zzr+09: \citet{zzr+09}.}
  \\
  \hline \hline
  Glon & Glat  &$V_{m}$ & $V_{p}$ & Ref. &  D & $\sigma_{D}$  & Ref.  & Mark & Ref.  & R$_{8.5}$ & D$_{8.5}$ & R$_{8.3}$  & D$_{8.3}$ & R$_{\rm BB93}$ & D$_{\rm BB93}$ \\
  \\
  $(^\circ)$& $(^\circ)$ &  km s$^{-1}$ &  km s$^{-1}$&   & kpc & kpc & & & &kpc &kpc &kpc & kpc & kpc & kpc\\
  (1) & (2) &(3) & (4) & (5)  & (6)  & (7) & (8) & (9) & (10) & (11) & (12) & (13) & (14) & (15) & (16)\\
  \hline
\endfirsthead
\caption{continued.}\\
\hline\hline
  Glon & Glat  &$V_{m}$ & $V_{p}$ & Ref. &  D & $\sigma_{D}$  & Ref.  & Mark & Ref.  & R$_{8.5}$ & D$_{8.5}$ & R$_{8.3}$  & D$_{8.3}$ & R$_{\rm BB93}$ & D$_{\rm BB93}$ \\
  \\
  $(^\circ)$& $(^\circ)$ &  km s$^{-1}$ &  km s$^{-1}$&   & kpc & kpc & & & &kpc &kpc &kpc & kpc & kpc & kpc\\
  (1) & (2) &(3) & (4) & (5)  & (6)  & (7) & (8) & (9) & (10) & (11) & (12) & (13) & (14) & (15) & (16)\\
\hline
\endhead
\hline
\endfoot
   0.212 &    0.000 &           45.5&   49.2&  pmb05&        &       &         &      &       ,17430$-$2844 pmb05            &      &       &       &      &       &       \\
   0.315 &   $-$0.200 &           20.5&   18.0&  pmb05&        &       &         &      &       ,17439$-$2845 pmb05            &      &       &       &      &       &       \\
   0.340 &    0.000 &           37.5&   37.0&  pmb05&        &       &         &      &       ,17432$-$2835 pmb05            &      &       &       &      &       &       \\
   0.393 &   $-$0.034 &           26.5&   28.7&  pmb05&        &       &         &      &                                    &      &       &       &      &       &       \\
   0.496 &    0.188 &           $-$5.0&    0.8&  pmb05&        &       &         &      &       ,17429$-$2823 pmb05            &      &       &       &      &       &       \\
   0.530 &    0.180 &           $-$2.0&    7.0&  pmb05&        &       &         &      &       ,17430$-$2822 pmb05            &   &   &    &  &       &       \\
   0.546 &   $-$0.851 &           14.0&   14.0&  pmb05&        &       &         &      &       ,17470$-$2853 pmb05            &      &       &       &      &       &       \\
   ... &   ... &           ...&   ...&  ...&  ...      &  ...     &  ...       &  ...    &    ...            & ...     &  ...     &  ...     & ...     &  ...     & ...      \\

\hline\hline
\end{longtable}
\normalsize



\end{appendix}

\end{document}